\documentclass{jfm}
\pdfoutput=1 
\usepackage{graphicx}
\usepackage{mathtools}
\usepackage{units}
\usepackage{xcolor}
\usepackage{amssymb}
\usepackage{amsmath,mathrsfs}
\usepackage{amsmath,amssymb,amsfonts,bbm}
\usepackage[bbgreekl]{mathbbol}
\usepackage{mathrsfs}
 \usepackage{overpic} 
 \usepackage{hhline}
\usepackage{multirow}
\usepackage{array}

\definecolor{darkcyan}{rgb}{0.0, 0.55, 0.55}

\shorttitle{Flow past a sphere translating in a rotating fluid}
\title{Flow past a sphere {\color{black}{translating}} along the axis of a rotating fluid: Revisiting numerically Maxworthy's experiments}

\author{Tristan Auregan\aff{$\dagger$}, Thomas Bonometti and Jacques Magnaudet}

\affiliation{Institut de M\'ecanique des Fluides de Toulouse (IMFT), Universit\'e de Toulouse, CNRS, INPT, UPS, Toulouse, France \aff{$\dagger$} Present address: Univ. Paris, Sorbonne Univ., Univ. PSL, PMMH, CNRS, ESPCI Paris, F-75005 Paris, France}

\begin{document}
\maketitle

\abstract{
We compute the flow induced by the steady translation of a rigid sphere along the axis of a large cylindrical container filled with a low-viscosity fluid set in rigid-body rotation, the sphere being constrained to spin at the same rate as the undisturbed fluid. 
The parameter range covered by the simulations is similar to that explored experimentally by Maxworthy [\textit{J. Fluid Mech.}, vol. 40, pp. 453-479 (1970)]. We describe the salient features of the flow, especially the internal characteristics of the Taylor columns that form ahead of and behind the body and the inertial wave pattern, and determine the drag and torque acting on the sphere.  
Torque variations are found to obey two markedly different laws under rapid- and slow-rotation conditions, respectively. The corresponding scaling laws are predicted by examining the dominant balances governing the axial vorticity distribution in the body vicinity. Results for the drag agree well with the semi-empirical law proposed for inertialess regimes by Tanzosh \& Stone [\textit{J. Fluid Mech.}, vol. 275, pp. 225-256 (1994)]. This law is found to apply even in regimes where inertial effects are large, provided rotation effects are also large enough. 
Influence of axial confinement is shown to increase dramatically the drag in rapidly rotating configurations, and the container length has to be approximately a thousand times larger than the sphere for this influence to become negligibly small. 
The reported simulations establish that this confinement effect is at the origin of the long-standing discrepancy existing between Maxworthy's results and theoretical predictions. 

}

\section{Introduction}
\label{sec:intro}
The spectacular and subtle characteristics of the flow field generated by a rigid or deformable body translating in a rapidly rotating fluid have received much attention for more than a century, starting with the landmark investigations of \cite{Proudman1916} and \cite{Taylor1917}. This configuration, which shares similarities with flows in stratified or magnetized fluids, is of practical relevance in problems where particles, drops or bubbles settle or rise in locally rotating flows, such as, e.g., in the dynamics of rapidly rotating suspensions or in centrifugal separation techniques employed in two-phase flows \citep*{Ungarish_1993, bush1994particle}. It is also relevant in ocean and atmosphere dynamics \citep{Loper2001} and, combined with thermal or compositional convection, in astrophysics to understand the dynamics of liquid cores in terrestrial and rapidly rotating planets \citep*{Bush1992,Cheng2015}. 

The flow disturbance generated by a rigid axisymmetric body with equatorial radius $a$ moving at speed $U_{\infty}$ in a Newtonian fluid of kinematic viscosity $\nu$ rotating as a whole with an angular velocity $\Omega$ depends on the Taylor number $\mathcal{T}a\equiv \frac{\Omega a^2}{\nu}$ and the Rossby number $\mathcal{R}o \equiv \frac{U_{\infty}}{\Omega a}$ (or equivalently the Reynolds number $\mathcal{R}e \equiv \frac{U_{\infty}a}{\nu}= \mathcal{R}o\mathcal{T}a$). 
Pioneering experiments with a cylinder or a sphere translating in a viscous fluid set in rigid-body rotation were performed by Taylor, with the body translating either parallel to the rotation axis \citep{taylor_1922} or perpendicular to it \citep{taylor_experiments_1923}. These experiments revealed the existence of slender recirculating fluid regions, later referred to as Taylor columns, confined within a cylinder circumscribing the body and having their generators parallel to the rotation axis. 
Later, Maxworthy repeated Taylor's 1922 experiments with a sphere translating along the rotation axis over a broad range of $\mathcal{T}a$ at both low Reynolds number \citep[$\mathcal{R}e \lesssim 0.5$,][]{maxworthy_experimental_1965} and moderate-to-large Reynolds numbers \citep[$5\lesssim \mathcal{R}e \lesssim 100$,][]{maxworthy_observed_1968}, \citep[$3\lesssim \mathcal{R}e \lesssim 300$,][]{maxworthy_flow_1970}. He confirmed Taylor's observations regarding the typical features of the flow structure, and found that the drag force on the sphere is generally increased by the fluid rotation, this increase scaling linearly with the Taylor number once the drag force has been normalized by the Stokes drag.

\begin{figure}
    \centering
    \includegraphics[height = 0.6\linewidth]{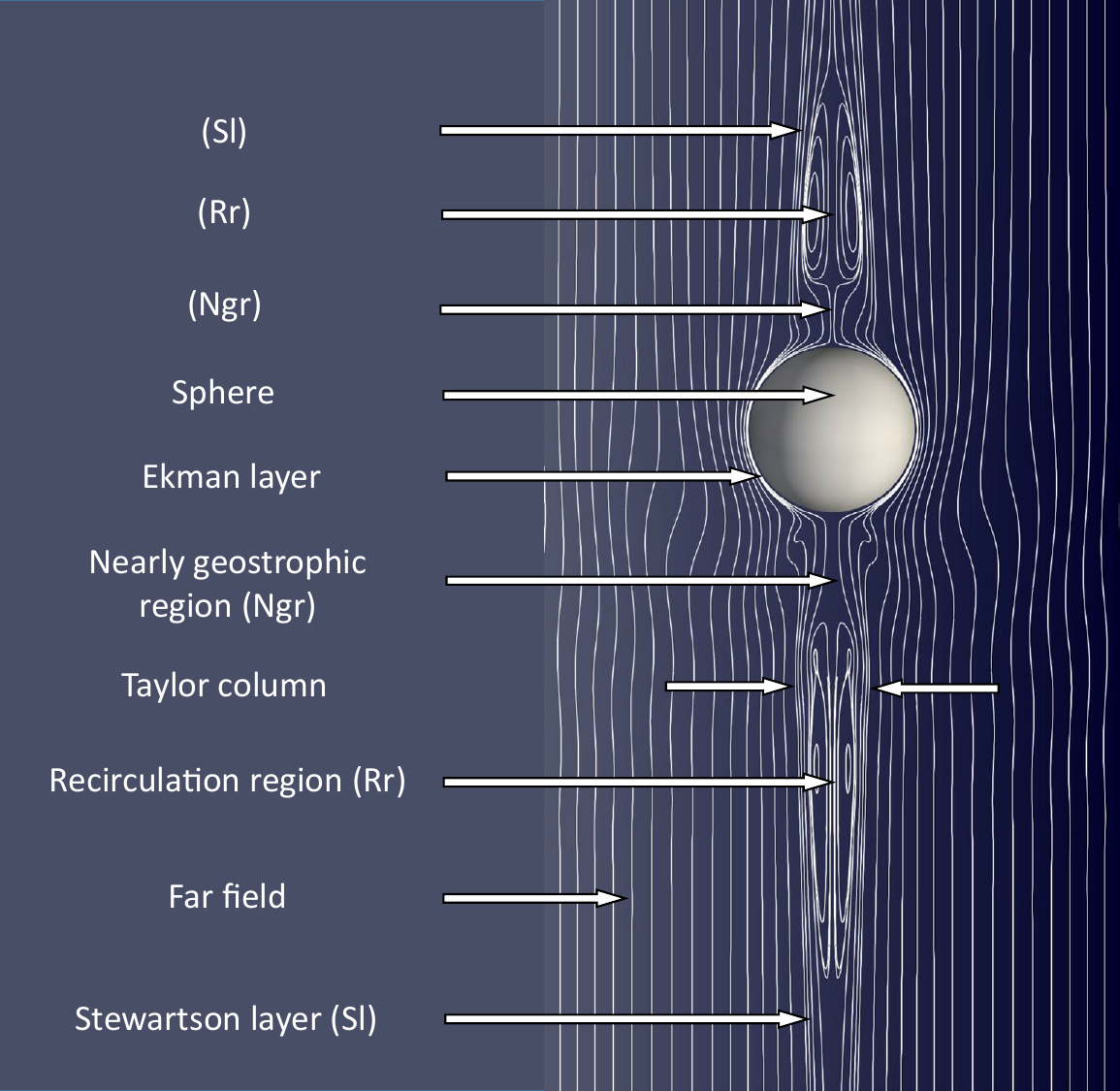}
    \caption{Qualitative flow structure past a sphere rising along the axis of a large fluid container set in rigid-body rotation. {\color{black}{The flow is observed in the reference frame translating with the sphere.}} The thin white lines are streamlines obtained from a simulation at $\mathcal{R}e=93$ and $\mathcal{T}a=193$, i.e. $\mathcal{R}o=0.48$.}
    \label{fig:flow_structure_new}
\end{figure}
A sketch of the corresponding flow at a relatively large Taylor number ($\mathcal{T}a=193$) is depicted in figure \ref{fig:flow_structure_new}. No fore-aft symmetry with respect to the sphere equator exists in this case, as advective effects are large ($\mathcal{R}e=93$). Two prominent recirculation regions standing upstream and downstream of the sphere may be observed. Existence of such recirculation regions in the present case is in line with the predictions of \cite{tanzosh_motion_1994} and \cite{vedensky_motion_1994}, the latter for a disc, which indicate that these structures take place when $\mathcal{T}a\gtrsim50$ and their axial extent (normalized by the body radius) grows approximately as $0.052\, \mathcal{T}a$. The second noticeable feature is the {\color{black}{nearly}} geostrophic region in which the Taylor-Proudman theorem {\color{black}{approximately}} applies \citep{moore_rise_1968}. In this smaller region, the non-dimensional length of which is approximately $0.006\, \mathcal{T}a$ \citep{tanzosh_motion_1994}, the fluid {\color{black}{almost}} achieves a rigid-body rotation, the rotation rate being faster (resp. slower) than $\Omega$ downstream (resp. upstream) of the body. {\color{black}{This nearly uniform swirling motion is accompanied by a weak plug-like axial flow thanks to which}} a tiny flux is transmitted from one {\color{black}{nearly}} geostrophic region to the other \textit{via} the Ekman boundary layer surrounding the body. The last salient flow feature is the Stewartson layer that connects the outer flow to the Taylor column (which is the body of fluid made of the recirculation and {\color{black}{nearly}} geostrophic regions and the above Ekman boundary layer). In the Stewartson layer, which has a complex internal  `sandwich' structure made of three concentric sublayers the thicknesses of which obey different scaling laws, an intense axial motion takes place while the swirl velocity varies rapidly in the radial direction \citep{Baker1967,moore_structure_1969}. This layer is the main region through which the fore and aft Taylor columns exchange fluid when the container is long enough for the end walls not to interact dynamically with these columns. \\
\indent Numerous studies have attempted to characterise the influence of the rigid-body rotation on the drag experienced by the sphere, both in finite-length and infinitely long containers.
\citet{stewartson_slow_1952} considered the asymptotic limit of an impulsive but slow motion in an \textit{inviscid} flow and an infinitely long container. Using a Laplace transform technique, he predicted that the drag force, $F_D$, is

\begin{equation}
    \frac{F_D}{F_{St}}=\frac{8}{9\pi}\mathcal{T}a \quad \Leftrightarrow \quad C_D=\frac{32}{3\pi}\mathcal{R}o^{-1}\approx 3.4 \mathcal{R}o^{-1} \quad (\mathcal{T}a= \infty,\,\mathcal{R}o\rightarrow0)\,,
\label{eq:stewartson}
\end{equation}
where $F_{St}=6\pi \rho\nu a U_{\infty}$ stands for the Stokes drag (with $\rho$ the fluid density), and the drag coefficient, $C_D$, is defined through the relation $F_D=\frac{1}{2}C_D\pi a^2\rho U_{\infty}^2 $. 
The above result was later confirmed by \cite{moore_structure_1969} assuming small-but-finite viscous effects. Conversely, \citet{childress_slow_1964} considered the \textit{viscous} regime and assumed $\mathcal{R}e\ll\mathcal{T}a^{1/2}\ll1$. Making use of the matching asymptotic expansion technique, he obtained

\begin{equation}
    \frac{F_D}{F_{St}}=1+\frac{4}{7}\mathcal{T}a^{1/2} \quad \Leftrightarrow \quad C_D=\frac{12}{\mathcal{R}e}+\frac{48}{7}(\mathcal{R}e\mathcal{R}o)^{-1/2} \quad (\mathcal{R}e\ll1, \mathcal{T}a\ll1,\,\mathcal{T}a/\mathcal{R}e^2\gg1)\,.
\label{eq:childress}
\end{equation}\noindent
Interestingly, Childress's theory also predicts that the drag is smaller than that in a non-rotating fluid when $\mathcal{T}a/\mathcal{R}e^2\lesssim0.2$, the largest reduction being $\approx5\%$ for $\mathcal{T}a/\mathcal{R}e^2\approx0.09$. Later, \citet{weisenborn_drag_1985} and \citet{tanzosh_motion_1994} predicted the drag for arbitrary Taylor numbers, still assuming the Reynolds number to be negligibly small. While both groups used distinct approaches (the so-called `induced-force' method and a boundary integral technique, respectively), the two sets of results are in agreement within 0.5\% up to $\mathcal{T}a=10^4$, and both agree within 5\% with the semi-empirical law proposed by \citet{tanzosh_motion_1994}, namely

\begin{equation}
    \frac{F_D}{F_{St}}=1+\frac{4}{7}\mathcal{T}a^{1/2}+\frac{8}{9\pi}\mathcal{T}a  \quad \Leftrightarrow \quad C_D=\frac{12}{\mathcal{R}e}+\frac{48}{7}(\mathcal{R}e\mathcal{R}o)^{-1/2}+\frac{32}{3\pi}\mathcal{R}o^{-1}  \quad (\mathcal{R}e \ll1)\,.
\label{eq:TS}
\end{equation}
The prediction \eqref{eq:TS} is nothing but the linear combination of \eqref{eq:stewartson} and \eqref{eq:childress}. Independently, \citet{vedensky_motion_1994} used a system of dual integral equations to predict the drag on a disc under similar conditions. {Effects of the finite length of the container were considered by \cite{moore_rise_1968}, assuming small-but-finite viscous effects and neglecting inertial effects. Considering a container with rigid ends and a half-length $H $ such that $1\ll\mathcal{L}\equiv H/a\ll\mathcal{T}a^{1/2}$, they showed that
\begin{equation}
    \frac{F_D}{F_{St}}=\frac{43}{630}\mathcal{T}a^{3/2} \quad \Leftrightarrow \quad C_D=\frac{86}{105}\mathcal{R}o^{-1}\mathcal{T}a^{1/2}  \quad (\mathcal{T}a \to \infty,\,\mathcal{R}o\rightarrow0)\,.
\label{eq:ms68}
\end{equation}
\textcolor{black}{Recently, \cite{kozlov2023motion} performed experiments with a sphere rising in a rapidly rotating short container ($\mathcal{L}=9.4$) in the range $\mathcal{T}a \in[250; 2.5\times10^{4}]$, $\mathcal{R}o \in [10^{-4}; 10^{-2}]$, and confirmed the $\mathcal{T}a^{3/2}$-dependence predicted by \eqref{eq:ms68}.} In this `short-container' limit, the Ekman layers that develop along the two end walls directly interact with the Taylor columns and ensure a good part of the fluid transport between the fore and aft columns, making the drag coefficient depend on viscosity (through the Taylor number), in contrast to  the `long-container' limit. In the latter, characterized by container aspect ratios such that $\mathcal{L}\gg\mathcal{T}a^{1/2}$, \textcolor{black}{the radial flow in these two Ekman layers is very weak and plays no role}. However, the end walls may still influence the internal structure of the Taylor columns through a purely kinematic `blocking' effect, thereby modifying the drag. For this reason, \cite*{Hocking1979} considered finite values of the ratio $\delta=\mathcal{L}/\mathcal{T}a$ (still in the limit on negligibly small Rossby numbers) and concluded that the drag increases monotonically as $\delta$ is reduced. For instance, when normalized by the prediction \eqref{eq:stewartson} corresponding to $\delta\rightarrow\infty$, they found that the drag on a sphere standing midway between the end walls increases by approximately $9\%$ for $\delta=1$ and $30\%$ for $\delta=1/4$.\\
\indent The low-Reynolds-number drag measurements ($\mathcal{R}e \lesssim 0.5$, $\mathcal{T}a \in [0.05, 0.7]$) carried out by  \citet{maxworthy_experimental_1965} agree within a few percent with \eqref{eq:childress}. It is worth noting that these data also support Childress' prediction that, at low enough $\mathcal{T}a/\mathcal{R}e^2$,  the drag is smaller than that in a non-rotating fluid. Conversely, at large enough Reynolds and Taylor numbers $(\mathcal{R}e \in [3; 300],\, \mathcal{T}a \in [10; 450])$, the data reported later by the same author \citep{maxworthy_flow_1970} follow the scaling \eqref{eq:stewartson}, albeit with a significantly larger pre-factor. Based on the comparison between \eqref{eq:stewartson} and \eqref{eq:ms68}, Maxworthy suspected that the origin of the discrepancy may stand in the finite length of his container, which was such that $\mathcal{L}\approx80$ or $\mathcal{L}\approx120$, depending on the size of the particles used. Hence, he corrected his results from end-wall effects using supplementary data, some of which, reported in \citet{maxworthy_observed_1968}, were obtained in a much shorter container ($5\lesssim\mathcal{L}\lesssim10$). Based on this correction, he concluded that his data may be extrapolated to an infinitely long container in the form
\begin{equation}
    C_D = (5.2\pm 0.1) \times \mathcal{R}o^{-1 \pm 0.01}\,.
\label{eq:max_extra}
\end{equation}
However, the pre-factor involved in \eqref{eq:max_extra} is still nearly $50\%$ larger than that in \eqref{eq:stewartson}. This discrepancy motivated the aforementioned extension of \eqref{eq:stewartson} to finite-length containers. However, the corresponding correction was found to only slightly reduce the discrepancy, making \cite{Hocking1979} conjecture that finite-$\mathcal{R}o$ effects not accounted for in their theory cannot be ignored.\\
\indent The very first simulations of the problem under consideration based on the full Navier-Stokes equations, hence incorporating finite-$\mathcal{R}o$ effects, were carried out by \cite*{dennis_slow_1982}. Computational limitations at that time restricted the explored parameter range to $\mathcal{R}e\le 0.5$ and $\mathcal{T}a\le0.5$. Nevertheless, these simulations were able to confirm quantitatively the experimental findings of \citet{maxworthy_experimental_1965} regarding the increase in drag with $\mathcal{T}a$ in the range $0\leq\mathcal{T}a\leq0.5$. \cite{Rao1995} explored a much broader range of Reynolds number (up to $\mathcal{R}e=500$) but only considered Rossby numbers larger than $2$. They could observe the changes in the flow structure in the presence of moderate rotation effects, especially the shrinking and disappearance of the standing eddy at the back of the sphere when $\mathcal{R}e\gtrsim20$ and $\mathcal{R}o$ is decreased from $\mathcal{O}(10)$ to $\mathcal{O}(1)$ values. They found that in this moderate-$\mathcal{R}o$, moderate-to-large-$\mathcal{R}e$ regime, rotation effects \textit{reduce} the drag, a finding also noticed by \cite{maxworthy_flow_1970} and later reconfirmed numerically by \cite{Sahoo2021}. \cite*{minkov_motion_2000, minkov_motion_2002} considered the case of a circular disc rising under low-$\mathcal{R}o$ conditions in short and long containers, respectively. They confirmed that the relative height of the container deeply affects the drag force. They also investigated the influence of the advective terms, i.e. finite-$\mathcal{R}o$ corrections, by exploring (in the long-container case) the range $\mathcal{R}o\le0.25$ with $\mathcal{T}a \in [100, 200]$, i.e. $\mathcal{R}e \leq50$. They concluded that these effects actually reduce the drag, thus further increasing the discrepancy with Maxworthy's data. \cite*{wang_numerical_2004} performed three-dimensional simulations of the same configuration for a sphere with or without a differential spin for $\mathcal{R}e=100$ and $250$ and $\mathcal{T}a \in [50, 6.25\times 10^3]$. They confirmed the characteristic features of the flow structure sketched in figure \ref{fig:flow_structure_new} at low Rossby number, and examined the influence of the control parameters on the inertial waves pattern. However, they did not report any drag value. Therefore, full Navier-Stokes simulations have not helped so far to reconcile the experimental findings of \cite{maxworthy_flow_1970} in the low-$\mathcal{R}o$ regime with theoretical predictions \eqref{eq:stewartson} or \eqref{eq:TS} for the drag. This is why the conclusion of \citet{minkov_motion_2002} that ``\textit{in any case, the major discrepancy between theory and experiments concerning
the value of the drag force remains unresolved, and becomes even more puzzling in view of the present results}'' still holds. 
\\
\indent This intriguing and unexplained discrepancy was the main initial motivation for the present work. We use fully resolved simulations to get new insight into this issue, and more generally into the influence of rigid-body rotation, viscous and advective effects on the organization of the flow past the body. The sphere is assumed to rotate at the same rate as the undisturbed flow and we determine the corresponding drag force and torque, assessing the possible influence of axial confinement effects on the flow structure and the loads on the body. We consider Taylor numbers $\mathcal{T}a \in [20, 450]$ and Reynolds numbers $\mathcal{R}e \in [2, 300]$, yielding Rossby numbers in the range $\mathcal{R}o \in [5\times10^{-3}, 10]$, which corresponds to the parameter range covered in Maxworthy's 1970 experiments. 
The mathematical problem, the numerical setup and a preliminary comparison with  zero-$\mathcal{R}o$ results are presented in \S\,\ref{sec:num_setup}. Characteristic features of the flow structure are discussed and compared with previous findings in \S\,\ref{sec:steady}. Then, the variations of the drag and torque with the control parameters are analysed in \S\,\ref{loads}. 
 The main outcomes of the study and some avenues for future work are presented in \S\,\ref{sec:concl}.

\section{Problem formulation and numerical setup}
\label{sec:num_setup}

\subsection{Governing equations and basic assumptions}
 \label{sec:equations}


 We assume that all flow characteristics are independent of the azimuthal position around the rotation axis, but the local velocity has a nonzero azimuthal component. 
\begin{figure}
    \centering
    \includegraphics[width = 0.6 \linewidth]{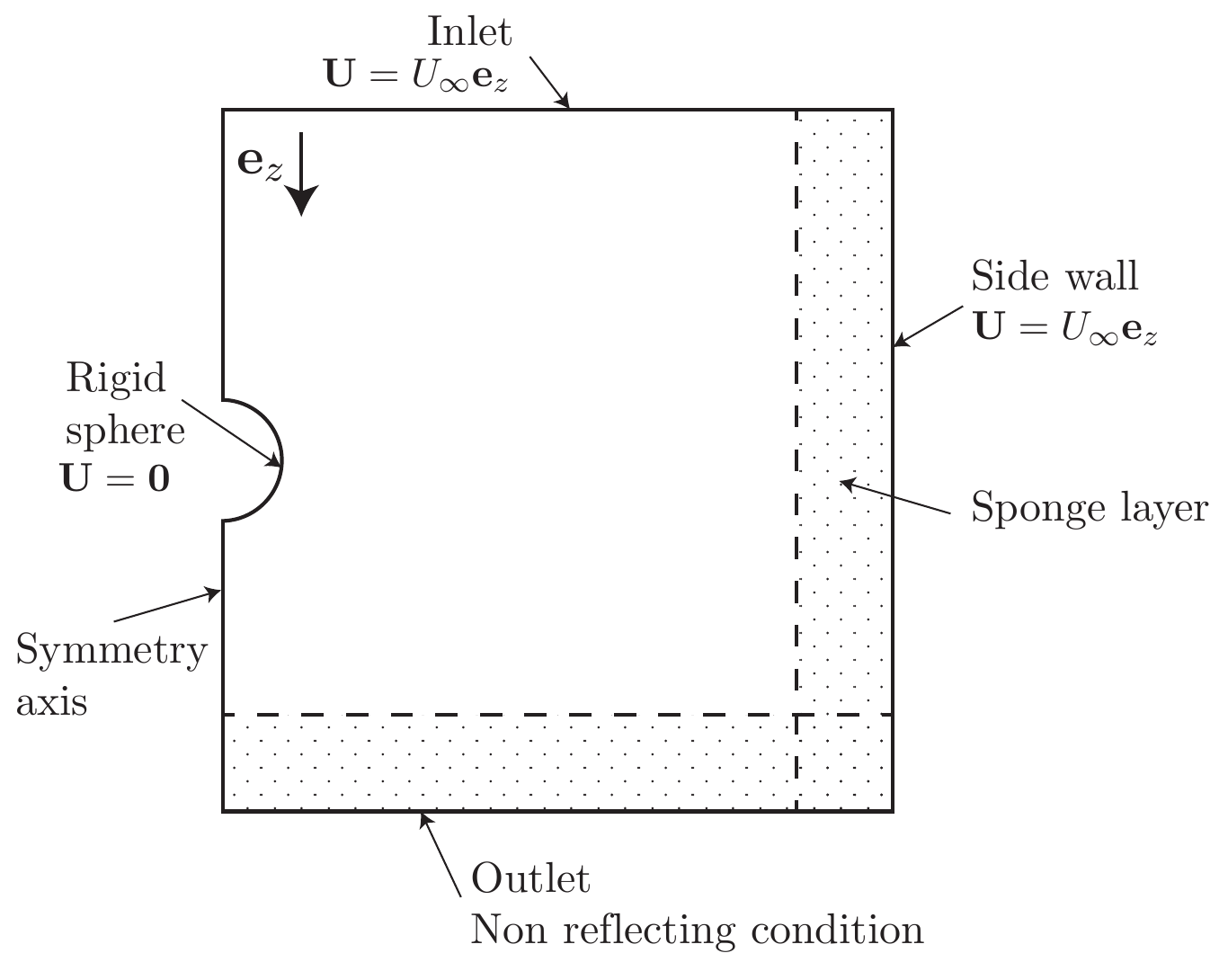}
    \caption{Sketch of the computational domain and boundary conditions (not to scale). 
    }
    \label{fig:num_set}
\end{figure}
We further assume that the sphere rotates with the prescribed angular velocity of the container, which lies along the $z$-axis. This assumption is rigorously satisfied when the flow exhibits a perfect fore-aft symmetry with respect to the sphere equator, which is achieved in the limit $\mathcal{R}o=0$. Nevertheless, we also carried out additional computations covering the whole range of flow conditions of interest here with the torque-free condition. In \S\,\ref{torque}, it will be shown that switching from one condition to the other has a negligible influence on the drag as long as $\mathcal{R}o\lesssim1$, and only a modest influence at higher $\mathcal{R}o$, yielding relative drag differences of less than $10\%$. 
 Assuming the flow to be incompressible and the fluid to be Newtonian, with density $\rho$ and kinematic viscosity $\nu$, the {\color{black}{continuity and}} Navier-Stokes equations expressed in the reference frame rotating and translating with the sphere read
\label{NS}
\begin{equation}
\nabla \cdot \mathbf{U}=0\,;\quad\partial_t \mathbf{U}+\mathbf{U}\cdot\nabla \mathbf{U}
 = -\rho^{-1} \nabla P+ \nu\nabla \cdot  (\nabla\mathbf{U} + \nabla\mathbf{U}^\text{T}) -2 \mathbf{\Omega}\mathbf{e}_z\times \mathbf{U}\,,
\end{equation}

\noindent with $\mathbf{U}$ the velocity field, $P$ the modified pressure including the centrifugal contribution, $\Omega$ the imposed rotation rate and $\mathbf{e}_z$ the unit vector in the $z$-direction.

\subsection{Computational aspects} 
\label{sec:bc}
The computations are carried out with the second-order in-house finite volume code JADIM developed at IMFT. The spatial discretization of the velocity and pressure fields is performed on a staggered grid. 
Time integration of (\ref{NS}) is achieved by combining a third-order Runge-Kutta scheme for advective and Coriolis terms with a semi-implicit Crank-Nicolson scheme for viscous terms. 
Incompressibility is satisfied at the end of each time step through a projection method. The accuracy of the complete time-integration scheme is second order \citep{Calmet1997}. \\
\indent The boundary conditions are summarised in figure \ref{fig:num_set}. A uniform velocity $U_\infty \bf{e}_z$ is imposed on the upstream and lateral boundaries. 
{\color{black}{Since the reference frame translates and rotates with the sphere,}} the no-slip condition  
$\mathbf{U} = \mathbf{0}$ is enforced at the sphere surface, while on the flow axis the velocity components obey
\begin{equation}\label{eq:sym}
  {\color{black}{ { {\bf{e}}_z\times\bf{U}}={\bf{0}}\,,\quad \text{and} \quad {\bf{e}}_z\times\nabla({\bf{U}}\cdot{\bf{e}}_z)={\bf{0}}\,.}}
\end{equation}
 {\color{black}{Hence, only the axial velocity is nonzero on the axis and its normal derivative vanishes there.}} Last, the non-reflecting condition described by \cite*{magnaudet_accelerated_1995} is used on the downstream boundary. In short, the first (second) normal derivative of the tangential (normal)
velocity component is set to zero on this boundary, together with the second-order cross-derivative of the pressure. \\
\indent Since the axial sphere motion generates nonzero components of the Coriolis force, inertial waves take place when the Rossby number is small enough. These waves, whose wavelength is proportional to $U_\infty/\Omega$, are emitted by the sphere and propagate downstream and outwards. Therefore, they are not `naturally' evacuated from the computational domain. To prevent their energy from accumulating near the outer boundary, we add a sponge layer that progressively damps them without creating any reflection within the domain. 
For this purpose, we use the Rayleigh damping technique \citep{slinn_model_1998} which consists in replacing in this layer the exact velocity field $\mathbf{U}$ with the damped surrogate $\mathbf{U}^*$ defined as
\begin{equation}
    \label{eq:ray_damp}
   \mathbf{U}^*= \mathbf{U} - \alpha \left( \mathbf{U} - \mathbf{U}_0 \right),
\end{equation}
where $\mathbf{U}_0$ is some reference velocity reached by the flow close to the boundary, and $\alpha \in [0, 1]$ is a damping parameter. We select $\mathbf{U}_0 = U_{\infty} \mathbf{e}_z$ 
 and $ \alpha(\zeta) = \exp [ - 6.125 \left( \zeta / L_{sl} \right)^2 ]$, with $L_{sl}$ the thickness of the sponge layer and $\zeta$ the local distance from the relevant outer boundary. We choose $L_{sl}$ such that at least five cells stand in the sponge layer, which was found sufficient to damp efficiently the inertial waves while limiting the thickness of this specific region within which the numerical solution is unphysical. The quality of the solutions provided by the present code in association with the above sponge layer technique may be appreciated in the work of \cite*{zhang_core_2019} in the context of internal waves radiated by a sphere settling in a stratified fluid. A sketch of the computational domain specifying the treatment applied to each boundary is shown in figure \ref{fig:num_set}. 




\begin{figure}
    \centering
    \includegraphics[width = 0.6\linewidth]{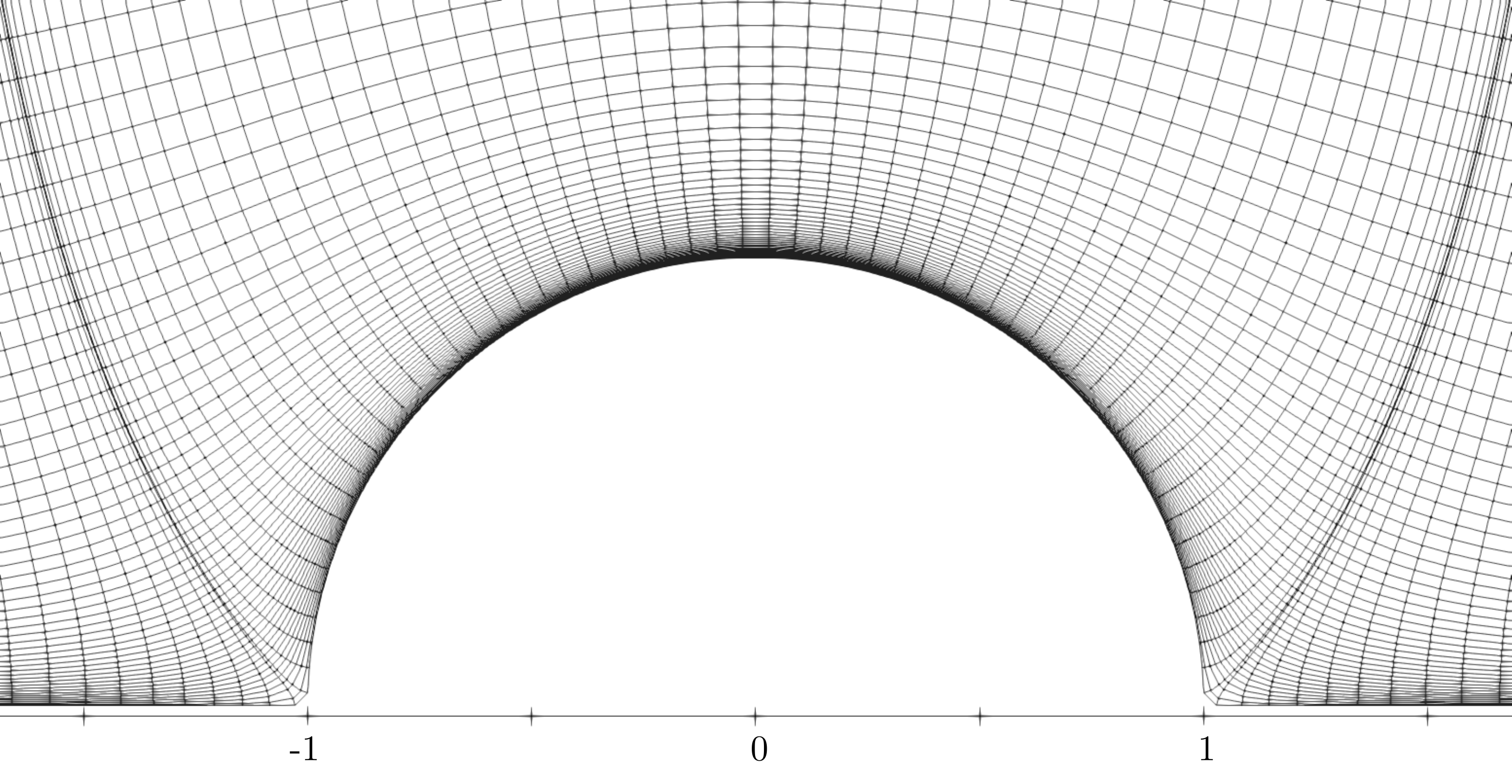}\\
     \includegraphics[width = 0.65\linewidth]{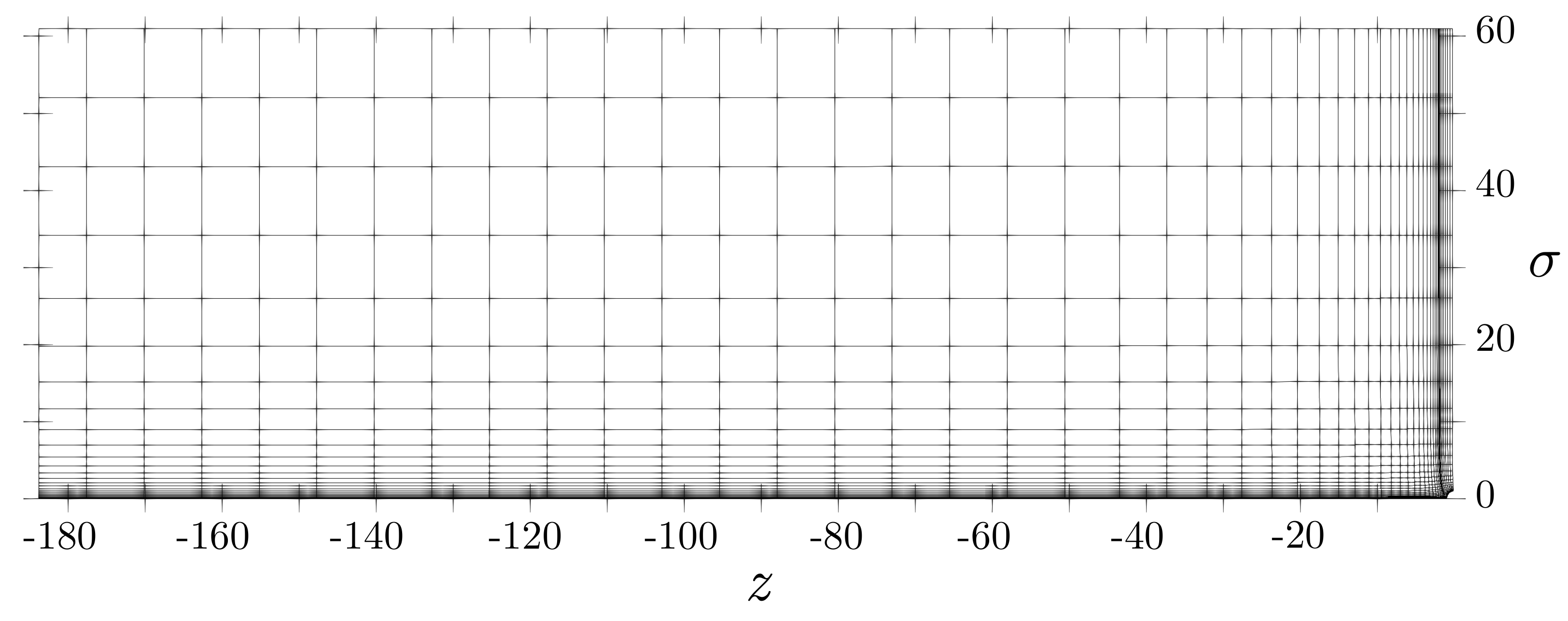}
    \caption{Computational grid (pressure nodes stand at the vertices). $(a)$: close-up view in the sphere vicinity; $(b)$upper semi-domain $z\leq0$ with $\mathcal{L}=180$ and $\mathcal{L}_\sigma=60$ (the sphere stands at the bottom right corner; 2 out of 3 cells have been removed in each direction for better visibility).}
    \label{fig:mesh}
\end{figure}

In JADIM, the Navier-Stokes equations (\ref{NS}) are expressed in a system of generalized orthogonal curvilinear coordinates. 
This makes it possible to use a variety of orthogonal boundary-fitted grids, as discussed by \cite{magnaudet_accelerated_1995}. {\color{black}{The detailed form of the governing equations expressed in this general coordinate system is also provided in this reference.}} Examples of solutions produced by this code associated with boundary-fitted grids for flows past spherical or spheroidal bodies, including in transitional or unstable regimes, may be found for instance in the works of \cite{Magnaudet2007} and \cite{Auguste2018}. Here, following \cite{magnaudet_accelerated_1995}, we employ an orthogonal grid built on the streamlines and iso-potential lines of the potential flow past a circular cylinder (figure \ref{fig:mesh}). {\color{black}{Accuracy of the solutions returned by the code on this type of grid may be appreciated in references such as \cite{magnaudet_accelerated_1995}, \cite{Legendre1998} and \cite{Legendre2003}, in which predictions for the forces acting on a spherical bubble in various two- and three-dimensional flow configurations are shown to compare very well with theoretical predictions in the limits of both low and high Reynolds number. \\
\indent With the above choice, the grid is nearly spherical in the sphere's vicinity (except close to the poles) and becomes gradually cylindrical as the distance to the sphere centre increases. }}Such a grid is particularly suitable for capturing efficiently not only the boundary layer surrounding the sphere, but also the wake and the near-axis upstream region even at very large distances from the body. {\color{black}{As will become apparent later, such far-field regions are of particular importance in the present problem and could hardly be captured with a spherical grid.}} 
The grid is non-uniform close to the sphere and becomes uniform far from it. Uniformity in the far field allows the thickness of the sponge layer to be properly controlled. The use of very thin cells along the sphere surface and the slow geometrical increase of the cell thickness as the distance to the sphere increases allow the `inertial' boundary layer (whose dimensionless thickness scales as $\mathcal{R}e^{-1/2}$) and/or the Ekman boundary layer (scaling as $\mathcal{T}a^{-1/2}$) to be properly captured throughout the considered range of parameters. Details on the grid design and a sensitivity study to some of the grid parameters are reported in appendix \ref{sec:conv_mesh}. When not stated otherwise, the half-length and radius of the computational domain (measured from the sphere centre and normalized by the sphere radius $a$) are $\mathcal{L} \times \mathcal{L}_\sigma=180 \times 60$, respectively, and the spatial discretization makes use of $314\times96$ cells. Nevertheless, following the discussion of \S\,\ref{sec:intro}, a detailed assessment of the influence of the axial confinement on the flow characteristics and the drag force is carried out in appendix \ref{sec:app_conf}, with $\mathcal{L}$ varied from $40$ to $\approx10^3$. Results of this sensitivity study are used in \S\S\,\ref{field} and \ref{loads} in the low-$\mathcal{R}o$ regime. 
 {\color{black}{Since the flow is expected to be invariant along the azimuthal direction over most of the conditions considered in Maxworthy's experiments, we opted for an axisymmetric resolution. Obviously, this simplification makes a parametric study much less expensive than a fully three-dimensional resolution. Nevertheless, it calls for some caution when the Reynolds number is large (typically $\mathcal{R}e\gtrsim100$) since the flow is known to be three-dimensional in that range in the absence of rotation. We shall come back to this issue at the beginning of \S\,\ref{field}.}} Starting from the uniform initial condition $\mathbf{U} = U_{\infty} \mathbf{e}_z$ throughout the flow domain, the computational time required to reach a converged stationary axisymmetric solution is approximately $2$ hours on a  standard single-processor workstation. The solution is considered converged when the relative time variation of the drag becomes less than $0.1 \%$ over $5\times10^4$ time steps. 
\subsection{Preliminary test} \label{sec:valid}

We first compare the local stress distribution at the sphere surface predicted with the above numerical setup at small-but-nonzero Reynolds number with those obtained by \citet{tanzosh_motion_1994} who made use of a boundary integral method in the creeping-flow limit. For this purpose we define the {\color{black}{stress tensor $ \mathbf{T}=-P\mathbf{I}+ \rho\nu \left (\mathbf{\nabla U} +  \mathbf{\nabla U}^{\text{T}} \right) $ ($\mathbf{I}$ denoting the Kronecker delta), and the}} surface traction $ \mathbf{n}\cdot\mathbf{T}\big|_{r=a} =F_r{\bf{e}}_r+F_\theta{\bf{e}}_\theta+F_\phi{\bf{e}}_\phi$, with  $\mathbf{n}\equiv{\bf{e}}_r$ the unit normal to the sphere pointing into the fluid, and 
$({\bf{e}}_r,\,{\bf{e}}_\theta, {\bf{e}}_\phi)$ the radial, polar and azimuthal unit vectors corresponding to the $(r,\theta, \phi)$ {spherical} coordinate system, with $r=0$ at the sphere centre and $\theta=0$ (resp. $\pi$) at the upstream (resp. downstream) pole. The $\theta$-variations of the three components of  the surface traction are displayed in figure \ref{fig:compts}.
The agreement is very good for the two tangential components, $F_\theta$ and $F_\phi$, although the values of the Taylor number in present simulations slightly differ from those of \citet{tanzosh_motion_1994}. 
The $F_r$-distributions also look similar but differences growing from the equator to the poles and reaching approximately $10\%$ close to the latter may be noticed. In particular, while the $F_r$-distributions reported by \citet{tanzosh_motion_1994} display an exact fore-aft antisymmetry (imposed by the $\mathcal{R}o=0$ assumption), those provided by present results do not. This is obviously due to finite Reynolds number effects. The reason why these effects manifest themselves essentially on $F_r$ is because this component of the traction reduces to the surface pressure, since the normal viscous stress vanishes on the sphere surface, owing to the combination of continuity and no-slip conditions. In contrast, only viscous stresses are involved in $F_\theta$ and $F_\phi$. Therefore, these traction components are less directly influenced by finite-$\mathcal{R}e$ effects, although a slight fore-aft asymmetry may be noticed in the central part of the distributions corresponding to $\mathcal{R}e=5$, most notably on $F_\phi$.  

\begin{figure}
    \centering
    \includegraphics[width = \linewidth]{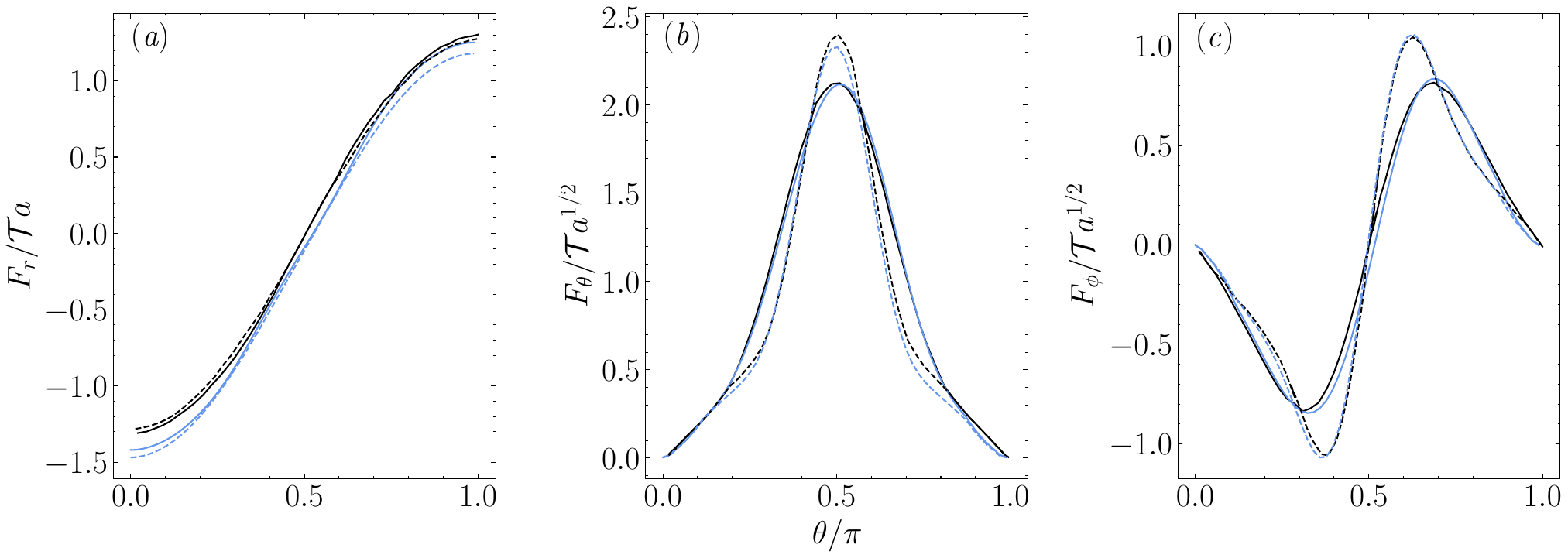}
    \caption{Variations of the three components of the surface traction, normalized by $\rho\nu U_{\infty} /a$, vs the polar angle $\theta$. Blue solid and dashed lines: present results for $(\mathcal{R}e = 5, \mathcal{T}a = 55.5)$ and $(\mathcal{R}e = 2, \mathcal{T}a = 445)$, respectively; black solid and dashed lines : zero-$\mathcal{R}o$ results of \citet{tanzosh_motion_1994} for $\mathcal{T}a=50$ and $\mathcal{T}a=500$, respectively. 
    }
    \label{fig:compts}
\end{figure}

\section{Flow field} \label{sec:steady}
\label{field}
\subsection{Preliminary comments}
\label{prelim}
We now examine the salient features of the flow fields provided by the simulations in the parameter range $\mathcal{T}a \in [20, 450]$, $\mathcal{R}o \in [10^{-2}, 10]$ (which makes the Reynolds number vary in the range $\mathcal{R}e \in [5, 300]$). By covering this range, we are in position to compare numerical predictions with the full set of experimental data reported by \cite{maxworthy_flow_1970}. 
However, it must be stressed again that present results were all obtained in axisymmetric simulations, although it is known that for large enough Rossby numbers the flow is already three-dimensional at Reynolds numbers less than the upper limit considered here. Indeed, in the absence of rotation ($\mathcal{R}o\rightarrow\infty$), it is established that axial symmetry in the wake of a translating sphere breaks down at $\mathcal{R}e\approx105$ \citep{Natarajan1993, Johnson1999}. In the presence of moderate rotation effects ($\mathcal{R}o=2$), \citet{wang_numerical_2004} showed that the flow past the sphere is still axisymmetric at $\mathcal{R}e=100$ but is three-dimensional and unsteady at $\mathcal{R}e=250$. {\color{black}{
However, since the governing equation for the vorticity becomes linear in the limit $\mathcal{R}o\rightarrow0$ (see the explicit form of the $z$-component of this equation below), no wake instability, hence no vortex shedding, can take place in this limit no matter how large the Reynolds number is.}} Consequently, it is expected that the lower $\mathcal{R}o$ is, the higher the critical Reynolds number for the onset of three-dimensional effects becomes.} A closely related increase of the critical $\mathcal{R}e$ below which the wake remains stable was reported by \cite{Machicoane2018} with a circular cylinder towed perpendicularly to the axis of a rapidly rotating container under conditions $\mathcal{R}o\lesssim10$. {\color{black}{More precisely, it was found that the cylinder's wake remains steady provided $\mathcal{R}e\lesssim550/\mathcal{R}o$, to be compared with $\mathcal{R}e\leq23.5$ in the limit $\mathcal{R}o\rightarrow\infty$. Hence, considering that the constraints imposed to the flow in the low-$\mathcal{R}o$ limit delay drastically the transition to three-dimensionality in the sphere's wake, we expect present axisymmetric results to remain valid up to the maximum considered Reynolds number ($\mathcal{R}e=300$) at low enough Rossby number, typically $\mathcal{R}o\lesssim1$ (unfortunately, how the critical $\mathcal{R}e$ varies precisely with $\mathcal{R}o$ is currently unknown). Results corresponding to Reynolds numbers larger than $105$ and $\mathcal{R}o\gtrsim1$ require some more caution. However, even in that range, the influence of three-dimensional, possibly unsteady, effects on the drag is still limited up to $\mathcal{R}e\approx200$. For instance, in a non-rotating flow, the time-averaged drag at $\mathcal{R}e=150$ is only $4\%$ larger than that predicted by constraining the flow to remain axisymmetric, and the relative amplitude of the drag oscillations is less than $1\%$ \citep{Tomboulides2000}. Consequently, the comparison of present predictions for the drag with experimental data in the same range (we carried out a series of runs at $\mathcal{R}e=167$) remains relevant. Only the few predictions corresponding to $\mathcal{R}e=300$ and $\mathcal{R}o >1$ may really suffer from the fact that three-dimensional effects, which yield a chaotic but not yet turbulent regime in the wake at this Reynolds number in a non-rotating flow \citep{Tomboulides2000, Poon2014}, are ignored in the present investigation.}}  
\subsection{General features} \label{sec:flow}

\begin{figure}
    \centering
    \includegraphics[height=0.92\textheight]{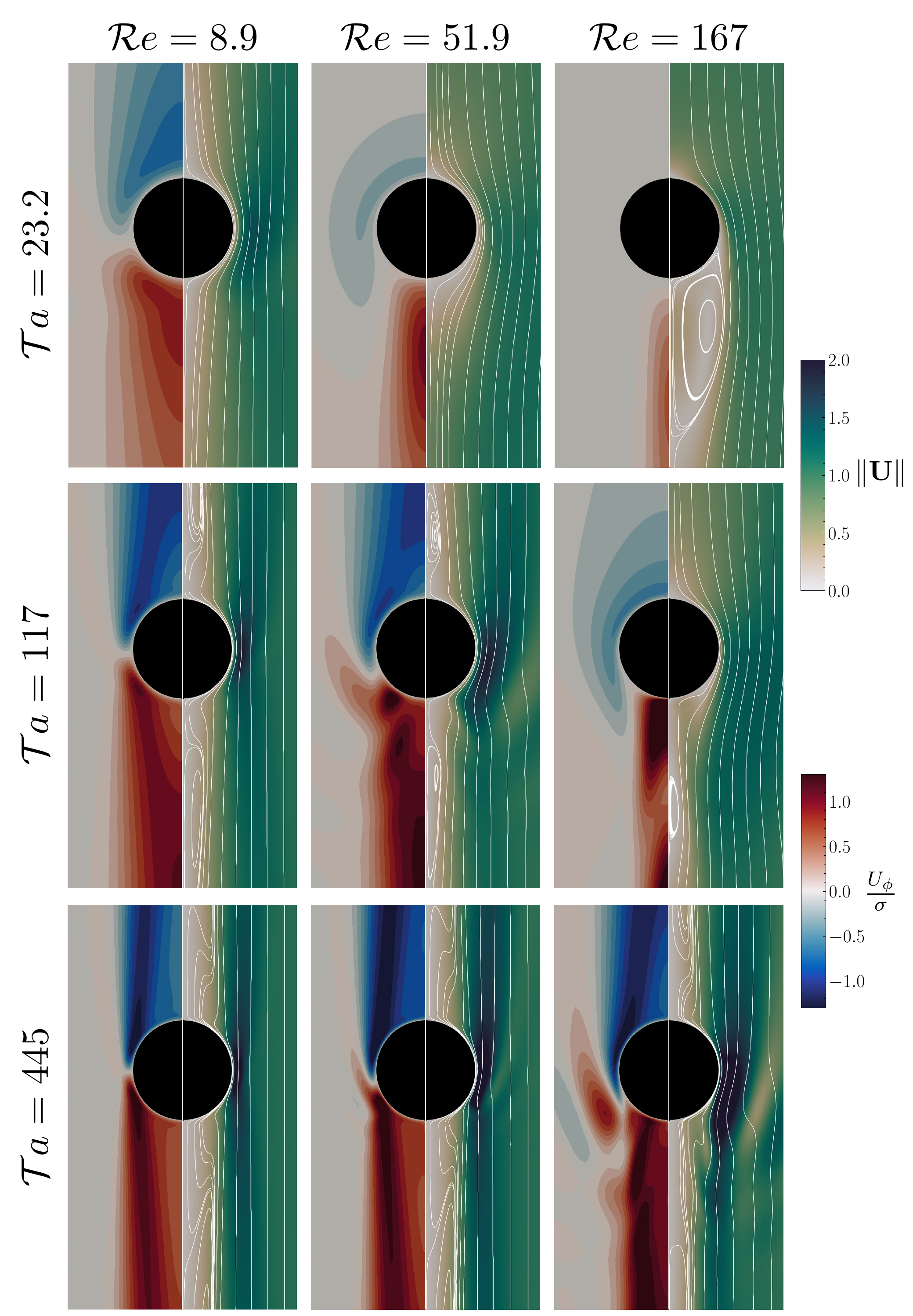}
    \caption{Flow structure past the sphere in the parameter space ($\mathcal{R}e, \mathcal{T}a$)
. The flow is from top to bottom. Left half of the panels (red-blue scale): angular velocity $U_{\phi} / \sigma$ (scaled by $U_\infty / a$); right half (scale of greens): velocity magnitude $||\bf{U}||$ (scaled by $U_{\infty}$) and streamlines in the sphere reference frame. }
    \label{fig:overview_iso}
\end{figure}
\begin{figure}
    \centering
    \includegraphics[height=0.92\textheight]{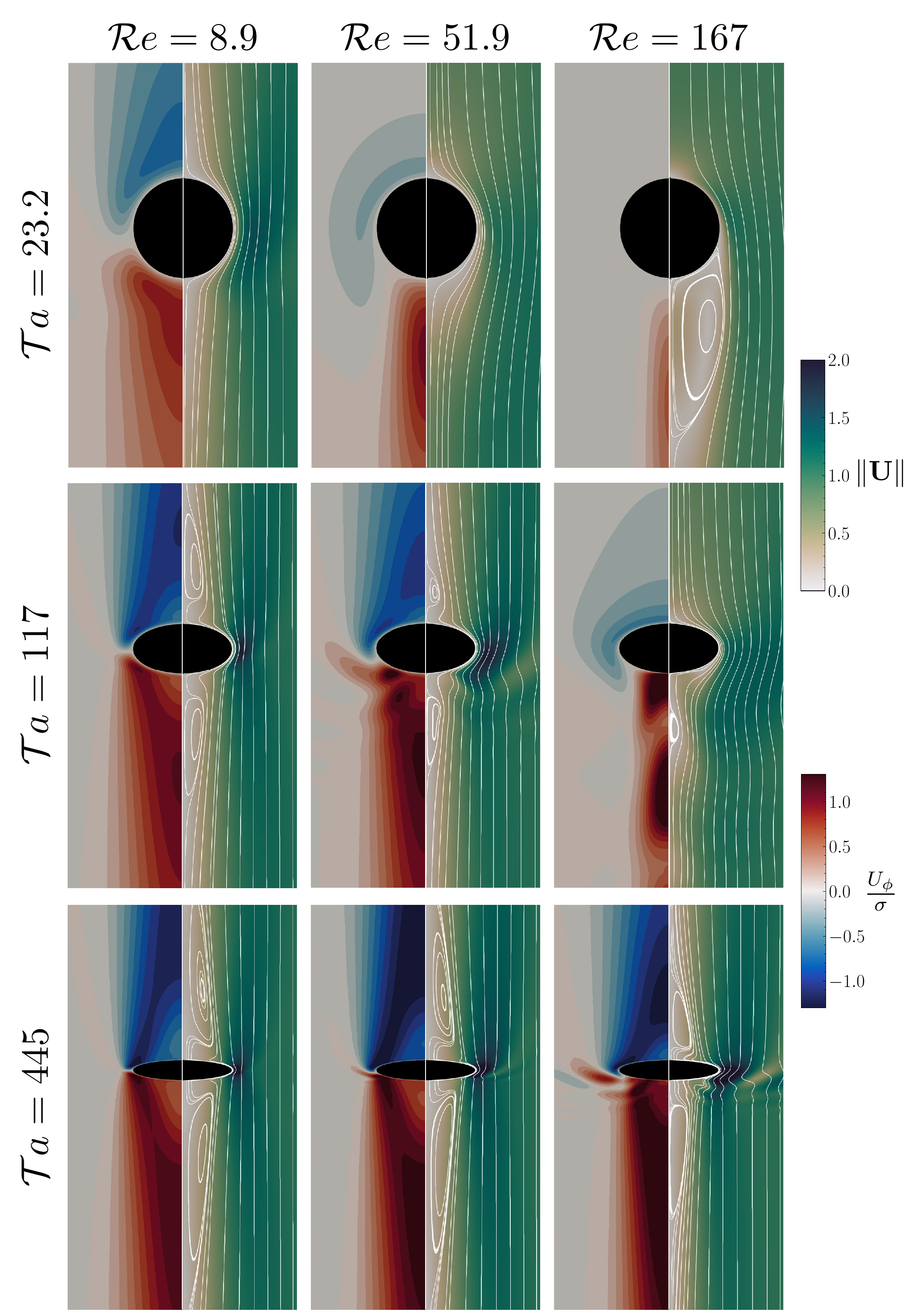}
    \caption{Same as figure \ref{fig:overview_iso} with the vertical axis compressed by a factor of $2$ (resp. $5$) for $\mathcal{T}a = 117$ (resp. $\mathcal{T}a = 445$) to capture the variations of the vertical extent of the recirculation regions. 
}
    \label{fig:overview}
\end{figure}
{\color{black}{From now on, we analyze the flow field using the cylindrical coordinates $\sigma, \phi, z$, with $\sigma$ the cylindrical radius ($\sigma=0$ on the rotation axis),  $\phi$ the azimuthal angle, and $z$ the axial distance from the sphere centre ($z<0$ upstream of the sphere, $z>0$ downstream of it).}} The flow past the sphere is presented in figures \ref{fig:overview_iso} and \ref{fig:overview} for various values of $\mathcal{T}a$ and $\mathcal{R}e$. Figure \ref{fig:overview_iso} allows to appreciate the details of the flow structure close to the sphere, while figure \ref{fig:overview} makes use of a compression along the vertical axis at the higher two $\mathcal{T}a$ to display the entire recirculation regions. 
Figure \ref{fig:overview_iso} evidences the vertical and radial growth of the Taylor columns as the Taylor number is increased and, for a given $\mathcal{T}a$, as $\mathcal{R}o$ is decreased by decreasing $\mathcal{R}e$. 
In line with previous observations, the fluid is seen to rotate more slowly (resp. faster) than the container in the upstream (resp. downstream) column. This feature may be rationalized by considering the governing equation for the axial vorticity, $\omega_z=\partial_\sigma U_\phi$, namely
\begin{equation}
\partial_t\omega_z+U_\sigma\partial_\sigma\omega_z+U_z\partial_z\omega_z-\omega_z\partial_zU_z-\omega_\sigma\partial_\sigma U_z=2\Omega\partial_zU_z+\nu\nabla^2_{\sigma,z}\omega_z\,,
\label{vortz}
\end{equation}
 where $\nabla^2_{\sigma,z}$ stands for the two-dimensional Laplacian operator {\color{black} {and $\partial_\sigma$ and $\partial_z$ denote the partial derivatives with respect to the cylindrical coordinates $\sigma$ and $z$, respectively}}. Noting that $\omega_\sigma=-\partial_zU_\phi$ and $\partial_zU_\phi\big|_{\sigma\ll a}\approx \sigma (\partial_z\omega_z)\big|_{\sigma=0}$, the vortex tilting term $-\omega_\sigma\partial_\sigma U_z$ may be approximated as $\sigma\partial_z\omega_z\big|_{\sigma=0}\partial_\sigma U_z$ near the rotation axis. Therefore all but one terms in \eqref{vortz} involve $\omega_z$ or its derivatives, which allows us to conclude that nonzero values of $\omega_z$ may only be created by the source term $2\Omega\partial_zU_z$. Moving towards positive $z$ along the generatrix $\sigma=a$, the no-slip condition at the sphere surface forces the flow to decelerate ahead of {\color{black}{the equatorial plane}}, implying $\partial_zU_z<0$ for $z<0$. Conversely, the flow must accelerate downstream of the {\color{black}{the equatorial plane}}, yielding $\partial_zU_z>0$ for $z>0$. Therefore, starting from rest, negative (resp. positive) values of $\omega_z$ are generated in the upper (resp. lower) part of the cylindrical region $\sigma\leq a$. {\color{black}{Normalizing velocities, distances, and time by $U_\infty, a$ and $a/U_\infty$, respectively, and denoting provisionally normalized quantities with an overbar, the non-dimensional form of \eqref{vortz} reads $\mathcal{R}o\,\overline{\textit{lhs}}=2\,\partial_{\overline{z}}\overline{U}_z+\mathcal{T}a^{-1}\nabla^2_{\overline{\sigma},\overline{z}}\overline{\omega}_z$, where $lhs$ stands for the left-hand side of \eqref{vortz}. 
 Now,}} the source term is of $\mathcal{O}(1)$, while the transport/stretching and diffusion terms are of $\mathcal{O}(\mathcal{R}o)$ and $\mathcal{O}(\mathcal{T}a^{-1})$, respectively. The steady-state distribution of $\overline{\omega}_z$ depends on the relative intensity of advection/stretching and viscous diffusion at a given $\mathcal{T}a$, hence on $\mathcal{R}o$ (or equivalently $\mathcal{R}e$). Considering frames $(a)-(c)$ for instance, the angular swirl $U_\phi/\sigma$ (which reduces to $\omega_z$ in the vicinity of the axis) is seen to approach a fore-aft symmetric distribution at $\mathcal{R}e=8.9$ (frame $(a)$), and to become increasingly asymmetric as the Reynolds number increases, with $\omega_z\approx 0$ upstream of the sphere at $\mathcal{R}e=167$ (frame $(c)$). In the latter case, advective effects are strong enough to reduce the flow region where $\omega_z$ exhibits significant values to a slender cylindrical zone in the wake. \\
\indent \citet{tanzosh_motion_1994} established that the recirculation regions appear at $\mathcal{T}a \approx 50$ in the zero-$\mathcal{R}o$ limit. Frames $(a)$ and $(d)$ in figure \ref{fig:overview}, which correspond to a fairly low Reynolds number, qualitatively support this prediction, as the former ($\mathcal{T}a =23.2$) reveals no recirculation while the latter ($\mathcal{T}a =117$) does. No upstream recirculation is found for $\mathcal{T}a = 117$ and $\mathcal{R}e=167$, i.e. $\mathcal{R}o=1.43$ (frame $(f)$), which suggests that the condition required for an upstream recirculation region to be present is actually $\mathcal{T}a \gtrsim 50$ \textit{and} $\mathcal{R}o\lesssim1$. 
Note that in the three panels corresponding to Rossby numbers larger than unity (frames $(b)$, $(c)$ and $(f)$), the spatial distribution of the angular velocity upstream of the sphere deeply differs from the columnar structure observed in all other cases. In frame $(f)$, the distribution downstream of the sphere looks also specific, with two well separated maxima located on both sides of a tiny standing eddy detached from the body.  
The flow structure in frame $(c)$ ($\mathcal{R}o=7.2$) is similar to that observed in a non-rotating case, with a large standing eddy extending downstream of the sphere. In contrast, no such structure is present in frame $(b)$ ($\mathcal{R}o=2.25$), indicating that rotation is now controlling the flow structure in the near wake. Therefore, it may be concluded that rotation effects start to manifest themselves when the Rossby number is below some units, typically $\mathcal{R}o\lesssim5$. A similar transition is observed with particles settling in a linearly stratified fluid, the Froude number based on the Brunt-V\"ais\"al\"a frequency then playing the role of the Rossby number \citep*{Torres2000,Magnaudet2020}. 
\begin{figure}
    \centering
    \includegraphics[width=0.99\linewidth]{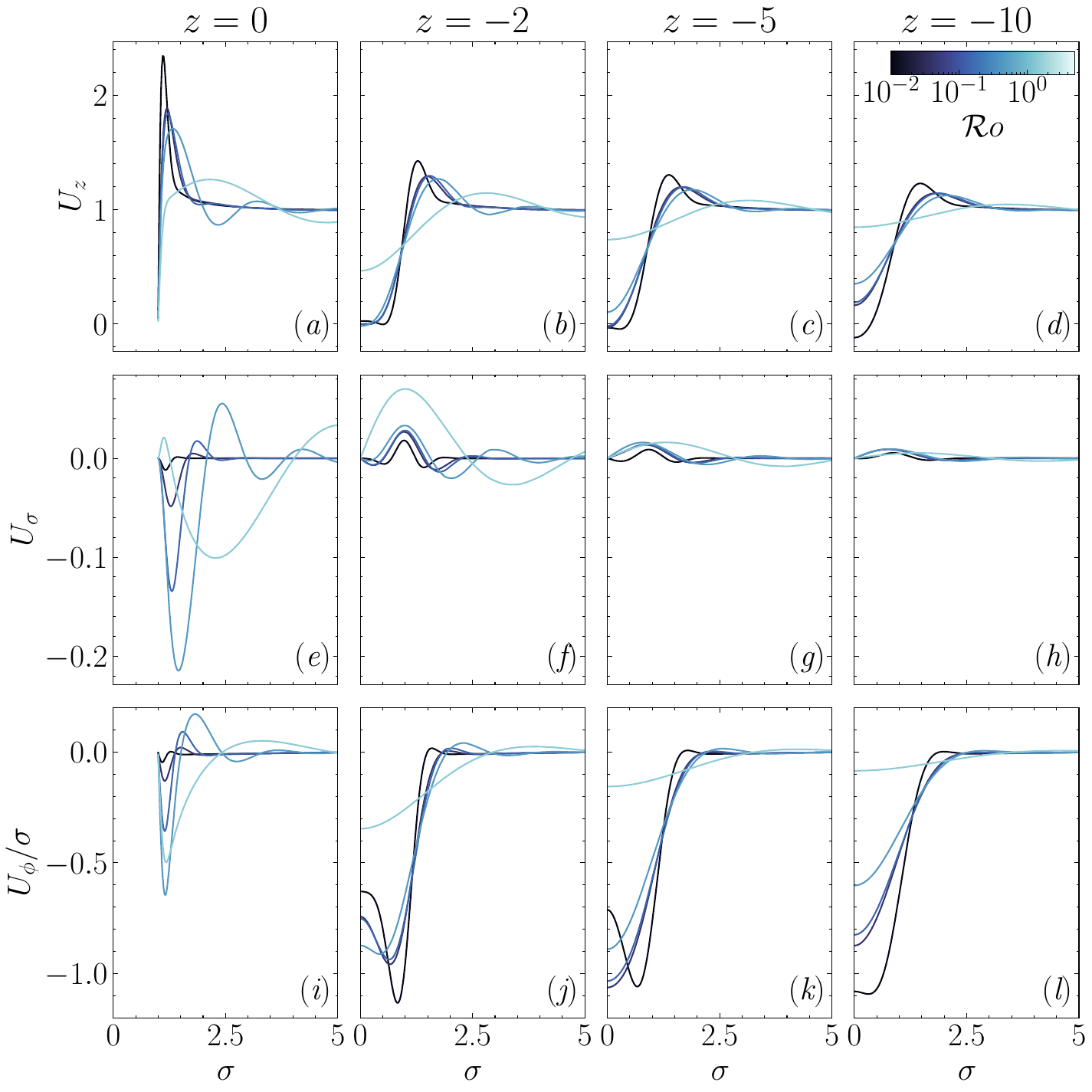}
    \caption{Radial slices of the velocity field in several $z=cst.$ planes ahead of the sphere: $(a)-(d)$: axial component  $U_z$; $(e)-(h)$: radial component $U_\sigma$; $(i)-(l)$: angular swirl $U_\phi/\sigma$. Velocities and distances are normalized by $U_\infty$ and $a$, respectively. Lines become lighter as the Rossby number increases. black: $\mathcal{R}o = 4.5\times10^{-3},\mathcal{R}e=2$ ($\mathcal{T}a = 445$); dark purple: $\mathcal{R}o = 4.3\times10^{-2},\mathcal{R}e=5$; dark blue: $\mathcal{R}o = 0.138,\mathcal{R}e=16.1$; medium blue: $\mathcal{R}o = 0.444,\mathcal{R}e=51.9$; pale blue: $\mathcal{R}o = 1.43,\mathcal{R}e=167$ ($\mathcal{T}a = 117$ in the latter four cases).}
    \label{fig:slices}
\end{figure}
When $\mathcal{R}o<1$, the flow becomes more and more one-dimensional as the Taylor number increases, with the Taylor column extending far upstream and downstream of the body (frames $(g-i)$). In the same frames, the radius of the upstream column is seen to decrease with the Rossby number, the Stewartson layer getting closer to the surface of the fluid cylinder circumscribing the sphere as $\mathcal{R}o\rightarrow0$. 
At the largest $\mathcal{T}a$, the upstream recirculation bubble extends more than 15 radii upstream of the sphere (frame $(g)$) and is shifted ahead of it by 2 radii. At such large $\mathcal{T}a$, the flow experiences strong variations along the sphere circumference, within the thin Ekman layer that surrounds {\color{black}{it}}. For instance, the fluid velocity near the equator is approximately $2.6$ times larger than the settling/rise velocity. 
\subsection{Near-body velocity distributions}
\indent Figure \ref{fig:slices} shows how the three velocity components vary under different flow conditions with the radial position in four successive planes perpendicular to the axis, from $z=0$ (equatorial plane) to $z=-10$, a plane standing within the recirculation region in the low-$\mathcal{R}o$ limit (lengths and velocities are considered dimensionless throughout this section, being normalized by $a$ and $U_\infty$, respectively). Disregarding provisionally the equatorial slice, one of the most significant features common to the three components is their large radial variation across the Stewartson layer standing around the mean position $\sigma=1$ and bounding externally the Taylor column. The peak values $U_z\approx1.4,\,U_\sigma\approx0.02,\,U_\phi/\sigma\approx-1.1$ reached by the three components in the plane $z=-2$ within this layer in the case $\mathcal{R}o=4.5\times10^{-3}$ agree well with the predictions of \cite{tanzosh_motion_1994} for $\mathcal{R}o=0$. Still with $\mathcal{R}o=4.5\times10^{-3}$, the near-axis plug-like profile of the axial velocity at $z=-2$ (frame $(b)$), with near-zero values up to $\sigma\approx0.6$, is typical of the {\color{black}{nearly}} geostrophic region. Moving upstream, $U_z$ is seen to take small negative values from the axis to $\sigma\approx0.4$ (frames $(c-d)$), which gives an estimate of the radius of the recirculation region. In contrast, $U_z$ keeps significant positive values whatever $z$ down to the axis in the most inertial case ($\mathcal{R}o=1.43$), which confirms the intuition that no  {\color{black}{nearly}} geostrophic or recirculation region exists under such conditions. Intermediate cases with $0.043\leq\mathcal{R}o\leq0.44$ (all with $\mathcal{T}a=117$) exhibit a nearly geostrophic behaviour up to $\sigma\approx0.3$ in the plane $z=-2$ (frame $(b)$). In contrast, the axial velocity keeps significant positive values down to the axis at $z=-10$ in these cases, showing that this plan stands beyond the tip of the recirculation region whatever the Rossby number for $\mathcal{T}a=\mathcal{O}(10^2)$.\\
\indent Returning to the case $\mathcal{R}o=4.5\times10^{-3}$, the near-axis profile of the angular swirl is seen to flatten gradually as the distance to the sphere increases, with on-axis values of $|U_\phi|/\sigma$ increasing from $0.6$ at $z=-2$ to $1.1$ at $z=-10$, approximately (frames $(j-l)$). Again, these findings are consistent with those of \cite{tanzosh_motion_1994}. Since $U_\phi/\sigma\approx\omega_z$ near the axis, the reason for this gradual increase and final plug-like profile may be understood by using \eqref{vortz}. When $\mathcal{R}o\rightarrow0$, axial variations of $\omega_z$ ahead of the sphere can only arise through the nonzero source term resulting from the weak axial variations of $U_z$. Radial variations of $\omega_z$ being negligible near the axis, one then has $2\mathcal{T}a\,\partial_zU_z\approx-\partial_{zz}\omega_z$. Thus, viscous diffusion is seen to induce a nonzero curvature in the axial profile of $\omega_z$. The axial velocity increasing from small negative values in the recirculation region to near-zero values in the  {\color{black}{nearly}} geostrophic region, the axial gradient $\partial_zU_z$ is positive, yielding $\partial_{zz}\omega_z<0$. Moreover, at a given radial location $\sigma\neq0$, $U_\phi$ increases from negative values upstream of the sphere to zero at its surface, while it remains null along the axis. Therefore, $\partial_\sigma(\partial_zU_\phi)$ is positive, implying $\partial_z\omega_z>0$ at the sphere surface. Combining the above two inequalities leads to the conclusion that $\partial_{z}\omega_z$ is necessarily positive (and larger than its surface value) ahead of the sphere, which translates into an increase of the angular swirl (in absolute value) as $|z|$ increases, in line with the behaviour observed in frames $(j-l)$. The argument still holds up to $z=-5$ for the two intermediate cases with $\mathcal{T}a=117$ and $\mathcal{R}o<0.2$. However, the plane $z=-10$ stands beyond the recirculation region in these cases, as the significant positive values of the axial velocity ($U_z\approx0.1$) confirm. Hence, $\partial_zU_z$ is negative and quite large beyond $z=-5$. This makes $\partial_{zz}\omega_z$ positive and significantly larger than in the zone closer to the sphere, leading to $\partial_{z}\omega_z<0$ beyond the recirculation region, and therefore to a reduction of the angular swirl as $|z|$ increases.  \\
\indent Symmetry arguments imply that the radial and azimuthal velocity components must both vanish on the equatorial plane at $\mathcal{R}o=0$. Hence, their nonzero values in that plane (frames $(e)$ and $(i)$) give insight into the strength of advective effects. Since these effects tend to enhance the amount of fluid transported from the upstream Taylor column to the downstream column through the Ekman layer, the main features of the $U_\sigma$ and $U_\phi$ near-surface distributions at $z=0$ when $\mathcal{R}o$ is nonzero are expected to resemble those found slightly above the equatorial plane in the zero-$\mathcal{R}o$ limit. The $\mathcal{O}(1)$ values of the axial velocity in the median part of the Stewartson layer upstream of the sphere (frame $(b)$), combined with its large peak values in the Ekman layer at $z=0$ (frame $(a)$), result in a positive $\partial_zU_z$ upstream of the equatorial plane at radial positions $\sigma\approx1$. Continuity combined with the no-slip condition at the sphere surface then implies $U_\sigma<0$ for $\sigma\gtrsim1$ above the equatorial plane. This is why, in the presence of finite inertial effects, one expects $U_\sigma$ to be negative near the sphere surface in that plane.
This is indeed the case as long as the near-surface peak of $U_z$ subsists  (frame $(e)$). More specifically, the magnitude of the (negative) peak value of $U_\sigma$ and $U_\phi/\sigma$ within the Ekman layer is seen to increase strongly with the Rossby number as long as $\mathcal{R}o$ is less than unity. The peak shifts away from the sphere surface as $\mathcal{R}o$ increases and its magnitude at $\mathcal{R}o=0.44$ is close to $0.2$ and $0.6$ for the (inward) radial velocity and angular swirl, respectively  (frames $(e)$ and $(i)$). The strongly inertial case corresponding to $\mathcal{R}o=1.43,\,\mathcal{R}e=167$ behaves differently, with especially $U_\sigma$ first taking positive values within the part of the boundary layer closest to the sphere surface  (frame $(e)$). Within the Ekman layer, the axial velocity reaches a maximum close to $2.4$ at $\mathcal{R}o=4.5\times10^{-3}$  (frame $(a)$). This value is in line with the findings of \citet{tanzosh_motion_1994} who reported a maximum of $2.25$ at $\mathcal{R}o=0$. The large positive values of $U_z$ within the Ekman layer play a pivotal role in the overall dynamics of the flow in the low-$\mathcal{R}o$ regime, as they directly control the amount of fluid transported from the upstream Taylor column to the downstream column. Inertial effects are found to change the $\mathcal{R}o=0$ picture dramatically, lowering the $U_z$ maximum to $1.7$ at $\mathcal{R}o=0.44$, which, taking the unit free stream velocity as reference, corresponds to a $50\%$ reduction of the peak. Putting the findings observed in the equatorial plane  on the three velocity components together, it appears that inertial effects deeply modify the local flow structure within the Ekman layer, which may be expected to have direct consequences on the stress distribution at the sphere surface, hence on the drag.

%

\subsection{Extent of the upstream recirculation region} \label{sec:recirc}

\begin{figure}
    \centering
    \includegraphics[width = 0.9\linewidth]{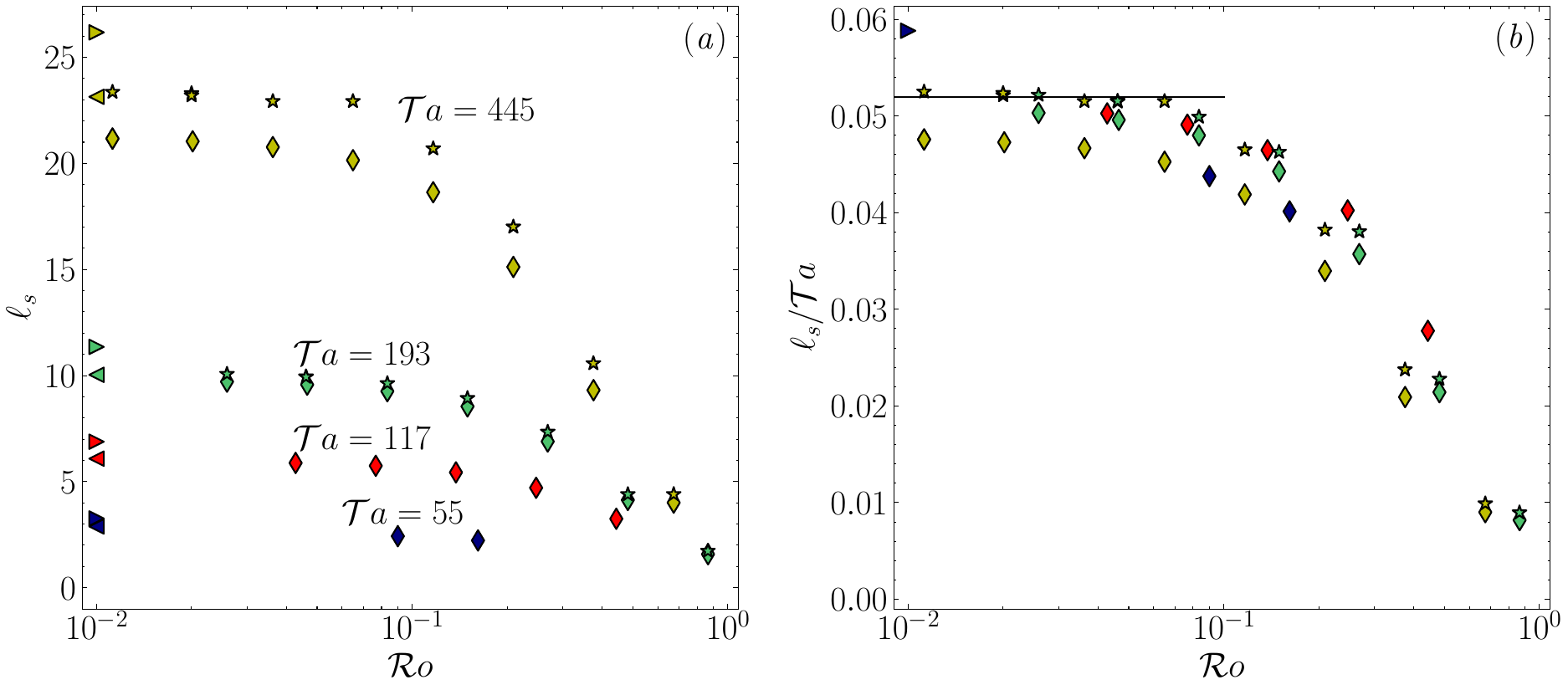}
    \caption{ Extent of the upstream recirculation region.  $(a)$: $\ell_{s}$  vs $\mathcal{R}o$ for various values of $\mathcal{T}a$; $(b)$: same with $\ell_{s}$ normalized by $\mathcal{T}a$. $\lozenge, \star$: present simulations with $\mathcal{L}=180$ and $\approx10^3$, respectively; $\triangleright$: experiments \citep{maxworthy_flow_1970}; $\triangleleft$: boundary-integral simulations at $\mathcal{R}o=0$  \citep{tanzosh_motion_1994}; \textcolor{black}{------} zero-$\mathcal{R}o$ limit $\ell_{s}=0.052\, \mathcal{T}a$ \citep{tanzosh_motion_1994}; dark blue, red, green and yellow symbols refer to $\mathcal{T}a=55,117,193$ and $445$, respectively. }
    \label{fig:lus_vs_ro}
\end{figure}


Figure \ref{fig:lus_vs_ro} shows how the extent of the upstream recirculation region, $\ell_{s}$, varies as a function of $\mathcal{R}o$ for various $\mathcal{T}a$. We define $\ell_{s}$ as the distance (normalized by the sphere radius $a$) from the sphere centre to the farthest upstream location where the axial velocity changes sign on the rotation axis. Strictly speaking, as figure \ref {fig:flow_structure_new} shows, the recirculation region stands in between the two locations where the axial velocity changes sign, the one closest to the sphere defining the tip of the \textcolor{black}{nearly} geostrophic region. However, we follow \citet{maxworthy_flow_1970} who, using dye visualisations, focused on the location of the tip of the recirculation region. In line with his observations, $\ell_{s}$ reaches a plateau  when $\mathcal{T}a$ is kept fixed and $\mathcal{R}o\rightarrow0$. Conversely, $\ell_{s}$ vanishes when $\mathcal{R}o \rightarrow 1$, implying that no recirculation region exists for $\mathcal{R}o>1$ (see also figures \ref{fig:overview} and \ref{fig:slices}). For a fixed $\mathcal{R}o$, $\ell_{s}$ decreases as $\mathcal{T}a$ is decreased, down to a critical Taylor number close to $50$ below which the recirculation region disappears. As figure \ref{fig:lus_vs_ro}$(b)$ shows, the simulations recover the prediction $\ell_{s} \approx 0.052\, \mathcal{T}a$ \citep{tanzosh_motion_1994} in the range $50 \lesssim \mathcal{T}a \lesssim 200$, $\mathcal{R}o \lesssim 5\times 10^{-2}$.
However, as the symbols in the upper left corner of figure \ref{fig:lus_vs_ro}$(a)$ reveal, the numerical results deviate from this prediction as well as from Maxworthy's data for the largest value of $\mathcal{T}a$ considered here, i.e. $\mathcal{T}a=445$. For instance, we find $\ell_{s}  \approx 21$ for $\mathcal{R}o = 2\times 10^{-2}$, which is significantly less than the value $\ell_{s}  =23$ reported by \citet{tanzosh_motion_1994} in the zero-$\mathcal{R}o$ limit. 
We attributed this discrepancy to axial confinement effects{\color{black}{, a track already suggested by \cite{ungarish_motion_1995}}}. To check this hypothesis, we increased the length of the computational domain from  $\mathcal{L}=180$ to  $\mathcal{L}\approx10^3$ along the lines discussed in appendix \ref{sec:app_conf}, where a detailed analysis of the sensitivity of the recirculation length and the drag force to these effects is presented. As the star symbols in figure \ref{fig:lus_vs_ro}$(b)$ show, the recirculation length obtained with this much longer domain is in excellent agreement with the zero-$\mathcal{R}o$ prediction. This is a clear indication that the characteristics of the recirculation region, and more generally those of the Taylor column, are extremely sensitive to axial confinement effects, even in containers with $\mathcal{L}=\mathcal{O}(10^2)$. Indeed, although the tip of the recirculation region stands far away from the top and bottom ends of the domain, the tip of the Taylor columns interacts directly with them when $\mathcal{T}a$ is large and $\mathcal{R}o\rightarrow0$, and the corresponding blocking effect is sufficient to alter the characteristics of the various zones of the flow located much closer to the body. 

\subsection{Inertial wave pattern}

\begin{figure}
    \centering
    \includegraphics[width=0.9\linewidth]{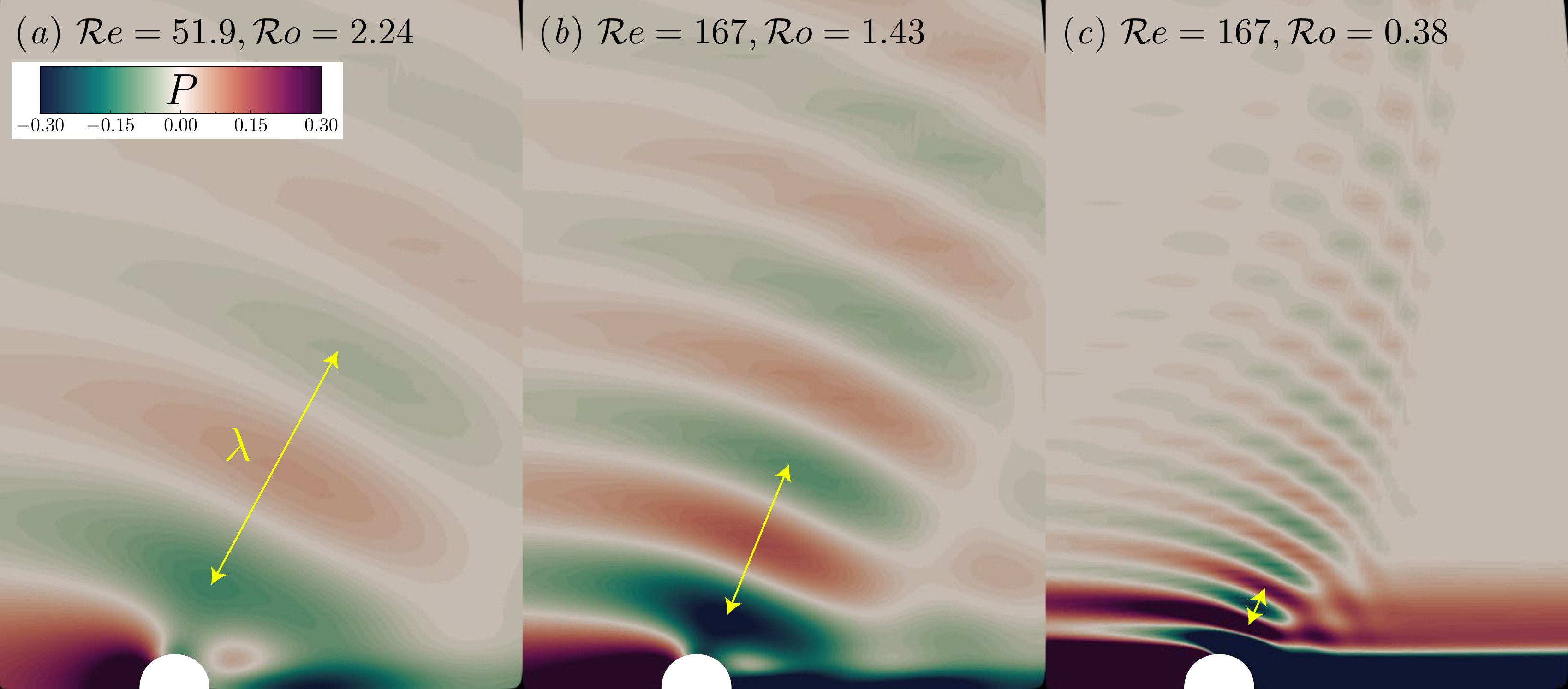}
    \caption{Visualization of the inertial wave pattern for several $\mathcal{R}o$; the {\color{black} {relative flow with respect to the sphere}} is from left to right. Colours refer to the amplitude of pressure variations (scaled by $\frac{1}{2}\rho U_\infty^2$). 
    }
    \label{fig:waves}
\end{figure}
\begin{figure}
    \centering
    \includegraphics[width=0.6\linewidth]{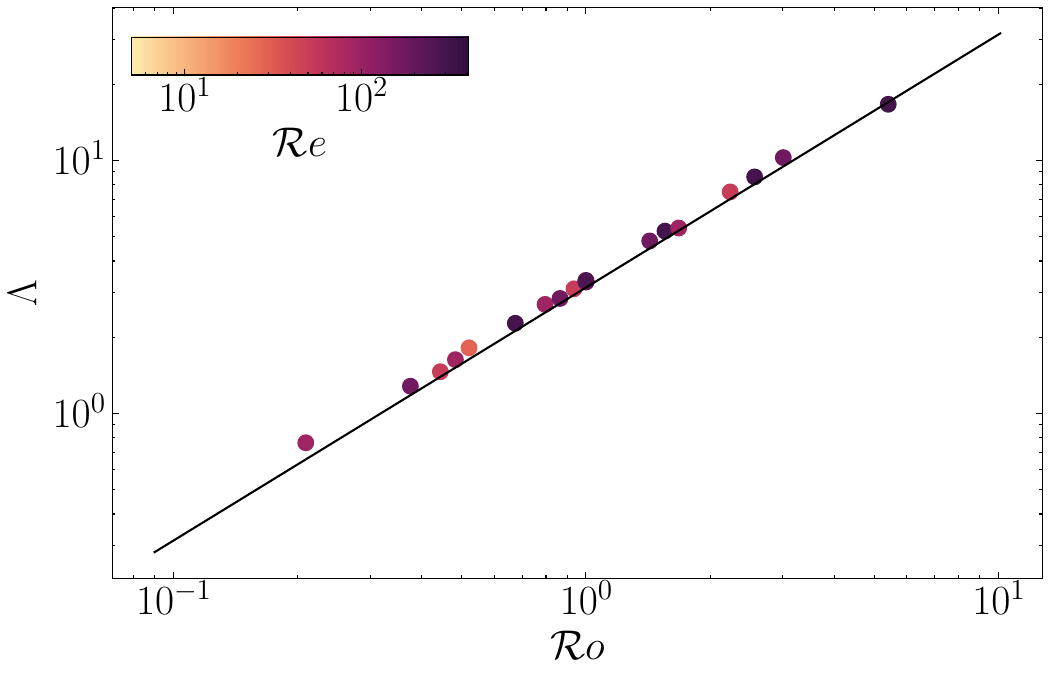}  
    \caption{ 
    Variations of the wavelength with respect to the Rossby number: \textcolor{black}{$\bullet$} simulations; \textcolor{black}{------} theoretical prediction $\Lambda=\pi \mathcal{R}o$. 
    }
    \label{fig:waves2}
\end{figure}

To finish with the characterization of the flow field, it is of interest to look at the dominant feature of the flow outside the Taylor column, namely the inertial wave field radiated by the sphere. The generation of such waves by bodies moving in a rotating fluid, or rotating topographies subject to a transverse flow, is well documented \citep{greenspan1968theory}. \citet{taylor_1922} predicted the existence of these waves and discovered that they exhibit an anisotropic dispersion property, with a radian frequency $\omega_{I0}$ obeying the orientation-dependent dispersion relation $\omega_{I0} = \pm2 \Omega \cos{\psi}$, with $\psi$ the angle between the wavevector ${\bf{k}}$ and the rotation axis. He also pointed out that, remarkably, this dispersion relation holds irrespective of the wave amplitude. However, unlike internal waves in a stably stratified fluid, pure inertial waves take place in a homogeneous fluid, which makes their experimental  observation more difficult \citep{Pritchard1969}. For this reason, Taylor could not observe the waves the existence of which he had predicted. Nevertheless, by releasing a light sphere on the axis of a rotating cylinder, he could visualise the existence of the resting column of fluid that was later named after him. Much later, waves generated by a pulsating, oscillating or transversely moving circular cylinder in a rotating tank could be visualised by \cite{Machicoane2015,Machicoane2018} using particle image velocimetry. We are not aware of similar experimental observations in the configuration considered here. The numerical investigation of \cite{wang_numerical_2004} provides some streamline maps for $\mathcal{R}o=2$ and $\mathcal{T}a=50$ and $125$, from which the wave pattern in a moderately rotating flow may be inferred.\\
\indent In the present configuration, the waves are radiated by the sphere moving relatively to the undisturbed fluid with velocity $-U_\infty\boldsymbol\Omega/\Omega$. Therefore, in the reference frame attached to the body, the radian frequency has to be corrected from the corresponding Doppler shift and becomes $\omega_I=\omega_{I0}+U_\infty{\bf{k}}\cdot\boldsymbol\Omega/\Omega$, 
so that $\omega_I$ obeys \citep{Lighthill1967,Whitham1974}
\begin{equation}
\omega_I= 2(\pi U_\infty/\lambda\pm\Omega) \cos{\psi}\,.
\label{disper}
\end{equation}
Equation (\ref{disper}) indicates that the relative displacement of the body along the rotation axis allows the existence of axisymmetric standing waves with wavelength $\lambda=\pi U_\infty/\Omega$, i.e. $\Lambda\equiv\lambda/a=\pi \mathcal{R}o$, as predicted by  \cite{taylor_1922}. 
In figure \ref{fig:waves}, we use the pressure disturbance field to visualise the wave pattern at three different values of the Rossby number. Figure \ref{fig:waves2}$(a)$ shows that the wavelength determined by seeking the minimum distance separating two successive crests (yellow arrows in figure \ref{fig:waves})  agrees closely with Taylor's theoretical prediction.The details of the wave field, i.e. the spatial distribution of $\psi$, are dictated by the no-penetration condition at the body surface and are influenced by viscous effects, especially those controlling the Ekman layer \citep{Johnson1982, Cheng1982}. \\
\indent Energy is radiated by the waves with the Doppler-shifted group velocity ${\bf{c}}_g=\partial_{\bf{k}}\omega_{I}={\bf{c}}_{g0}+U_\infty{\bf{e}}_z$ with ${\bf{c}}_{g0}=\partial_{\bf{k}}\omega_{I0}$. The axial and radial components of  ${\bf{c}}_g$ are $c_{gz}=U_\infty-\pi^{-1}\Omega\lambda\sin^2\psi$ and $c_{g\sigma}=(2\pi)^{-1}\Omega\lambda\sin2\psi$, respectively. Consequently, standing waves with $\lambda=\pi U_\infty/\Omega$ have axial and radial group velocities $c_{gz}=U_\infty\cos^2\psi$ and $c_{g\sigma}=\frac{1}{2}U_\infty\sin2\psi$, respectively, and their energy propagates along straight rays $\sigma=(z-z_0)\tan\psi$, with $z_0$ the origin of the ray on the rotation axis. 
According to figure \ref{fig:waves}, the angle $\psi$ increases from approximately $\pi/3$ downstream of the sphere to values close to $\pi/2$ upstream. Therefore, the axial and radial components of the energy flux are positive everywhere, i.e. they are directed downstream and outwards, respectively. Moreover, they decrease continuously as one moves upstream, and eventually vanish when the wave crests become parallel to the axis ($\psi=\pi/2$). Therefore, far upstream of the sphere, the wave energy does not propagate at all, i.e. it just travels with the sphere. Comparing the three subfigures indicates that the lower $\mathcal{R}o$ is the more the wave fronts are parallel to the rotation axis at a given position upstream of the sphere. Therefore, the lower $\mathcal{R}o$ the shorter the upstream position at which the wave energy stops propagating. 
\\
\indent Examination of the whole set of computational results reveals the presence of inertial waves with characteristics similar to those discussed above for Rossby numbers in the range $0.2 \lesssim \mathcal{R}o \lesssim 5$. These two limits result from totally distinct reasons. At $\mathcal{R}o= 5$, the wavelength is approximately one third of the radius of the computational domain (and even half that size if the sponge layer is not considered). Hence, the outer cylindrical boundary affects the distribution of the disturbances radiated by the body at larger $\mathcal{R}o$, preventing the formation of standing waves. Conversely, viscous effects are responsible for the disappearance of waves for $\mathcal{R}o\lesssim 0.2$. Indeed, a disturbance with wavevector ${\bf{k}}$ is damped at a rate $-\nu ||{\bf{k}}||^2$. Hence, the ratio $r_v$ between the viscous force and the restoring Coriolis force acting on the disturbance is $2\pi^2\nu/(\Omega\lambda^2)=2\pi^2/(\Lambda^2\mathcal{T}a)$, which for $\Lambda=\pi \mathcal{R}o$ yields $r_v=2(\mathcal{T}a\mathcal{R}o^2)^{-1}$. Consequently, the lower $\mathcal{R}o$ the larger $r_v$ at a given $\mathcal{T}a$, with for instance $r_v=1$ for $\mathcal{R}o=0.1$ and $\mathcal{T}a=200$.  

\section{Loads on the body}
\label{loads}
\subsection{Drag} \label{sec:drag}

\begin{figure}
    \centering
    \includegraphics[width = 0.8\linewidth]{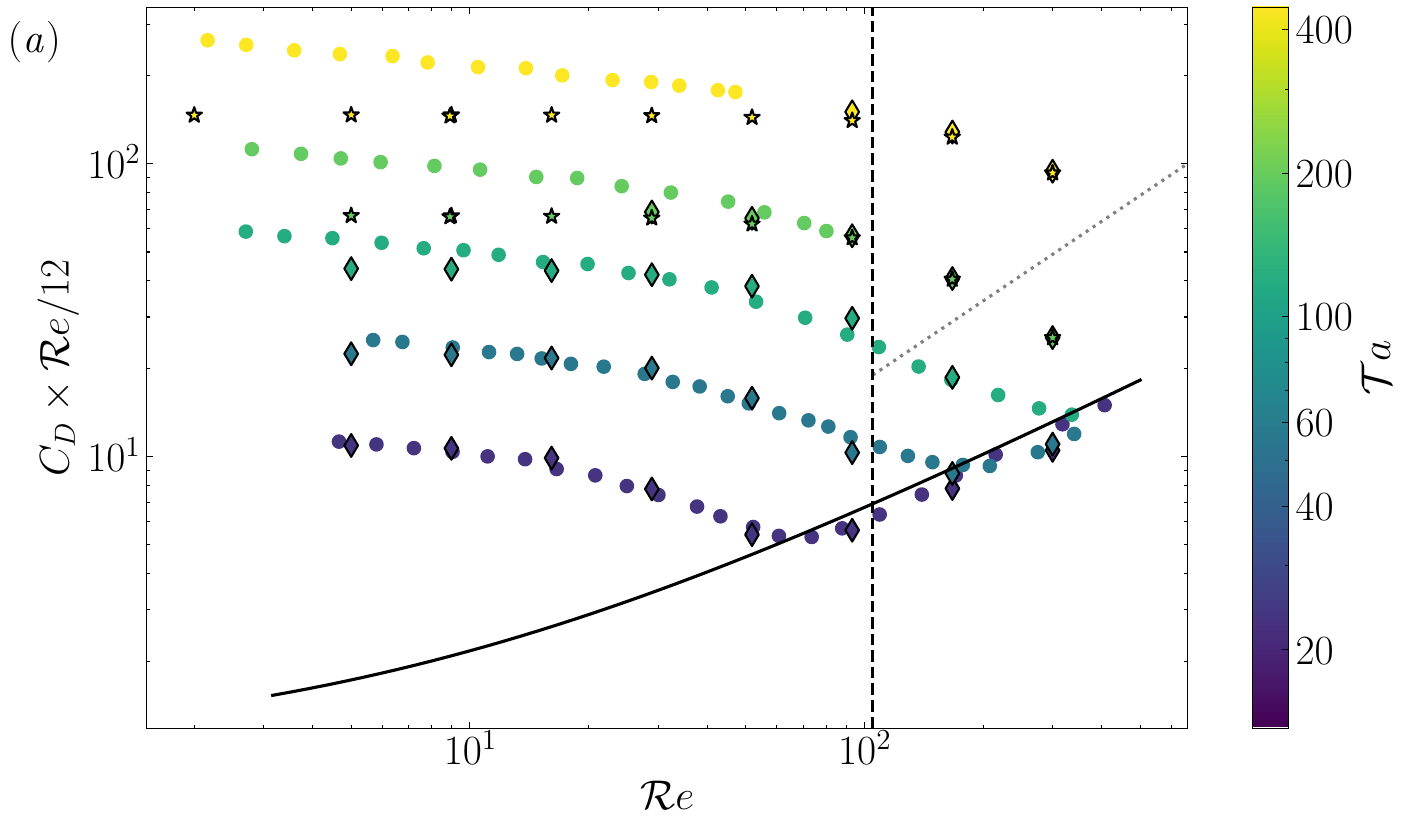} 
    \includegraphics[width = 0.8\linewidth]{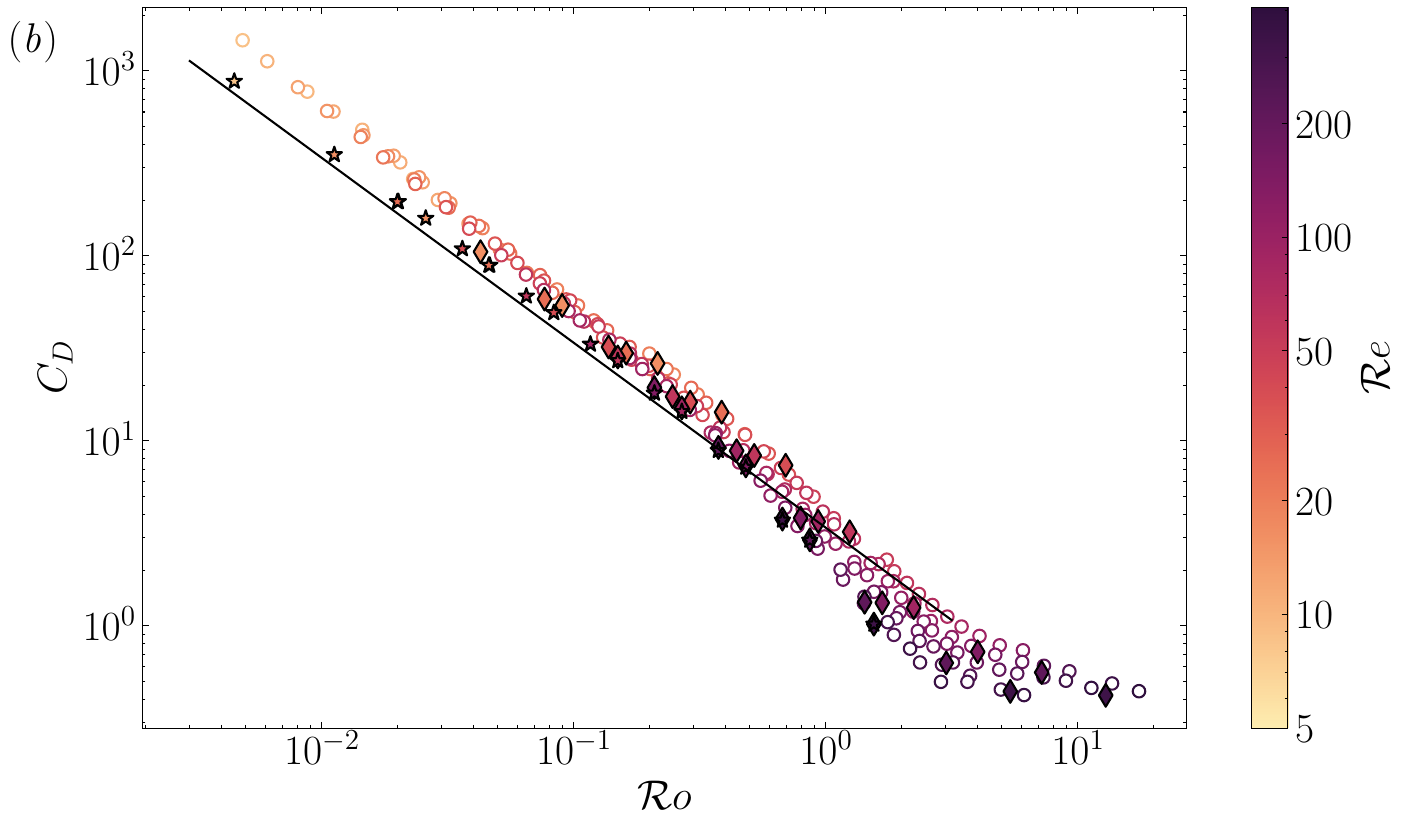}
    \caption{Drag coefficient vs  the Reynolds and Rossby numbers. $(a)$: $C_D$ vs $\mathcal{R}e$; $(b)$: $C_D$ vs $\mathcal{R}o$. $\blacklozenge, \bigstar$: present simulations with $\mathcal{L}=180$ and $\approx10^3$, respectively; $\circ$: experiments \citep{maxworthy_flow_1970}. In $(a)$, the solid line is the standard drag curve for a sphere translating in a fluid at rest; {\color{black}{the vertical dashed line corresponds to the threshold beyond which the wake is three-dimensional in the limit $\mathcal{R}o\rightarrow\infty$, and the dotted line is a guide for the eye separating data belonging to the range $\mathcal{R}o>1$ (below the line) from those for which $\mathcal{R}o<1$ (above the line).}} The solid line in $(b)$ corresponds to the inviscid prediction \eqref{eq:stewartson}. Symbols are colored according to the value of $\mathcal{T}a$ in $(a)$ and $\mathcal{R}e$ in $(b)$.  
 }
    \label{fig:d_vs_re}
\end{figure}

Figure \ref{fig:d_vs_re} presents the drag coefficient obtained through a direct integration of the surface traction defined in \S\,\ref{sec:valid} over the sphere. Subfigure $(a)$ shows the compensated drag coefficient $C_D\mathcal{R}e/12$ as a function of the Reynolds number for various $\mathcal{T}a$, while subfigure $(b)$ shows $C_D$ as a function of the Rossby number for various $\mathcal{R}e$. The standard drag curve for a sphere translating in a quiescent fluid, based on the empirical correlation $C_D(\mathcal{R}e)=12\mathcal{R}e^{-1}(1+0.241\mathcal{R}e^{0.687})$ \citep{Schiller1933}, and the inviscid prediction \eqref{eq:stewartson} are also shown as references. {\color{black}{For reasons discussed in \S\,\ref{prelim}, only the few numerical predictions corresponding to $\mathcal{R}e=300$ and $\mathcal{R}o>1$ (i.e. the three rightmost lozenges located below the dotted line in figure \ref{fig:d_vs_re}$(a)$) are expected to be significantly altered by the absence of three-dimensional effects in the computed solutions.}}

At low to moderate Reynolds number, say $\mathcal{R}e\lesssim50$, the drag is significantly larger than predicted by the above correlation, highlighting the influence of the rigid-body rotation. Present results agree well with those of  \cite{maxworthy_flow_1970} up to $\mathcal{T}a\approx80$, i.e. in the range where rotation effects are moderate. In contrast, they clearly deviate from experimental data for larger $\mathcal{T}a$, predicting a lower drag. The lower $\mathcal{R}e$ is, the larger the deviation at a given $\mathcal{T}a$ is, the relative difference between the two values exceeding $40\%$ at $\mathcal{R}e=5$ for the highest $\mathcal{T}a$. Conversely, the larger $\mathcal{T}a$, the larger the Reynolds number at which the deviation starts. Thus, numerical predictions and experimental data still agree for large enough $\mathcal{R}e$ when $\mathcal{T}a$ is large. 
In the $C_D$ vs $\mathcal{R}o$ representation of figure \ref{fig:d_vs_re}$(b)$, numerical predictions are seen to fall within the somewhat scattered interval of experimental values for $\mathcal{R}o \gtrsim 2\times10^{-1}$. 
In contrast, below this threshold, the numerical series departs from the experimental one, and the departure increases as $\mathcal{R}o$ decreases. It may be noticed that all numerical data obtained for $\mathcal{R}o\lesssim0.3$ stand beyond the inviscid prediction \eqref{eq:stewartson}. As these data correspond to Reynolds numbers less than $200$, viscous effects are likely to be responsible for the observed difference. This will be confirmed later.

\begin{figure}
    \centering
    \includegraphics[width=\textwidth]{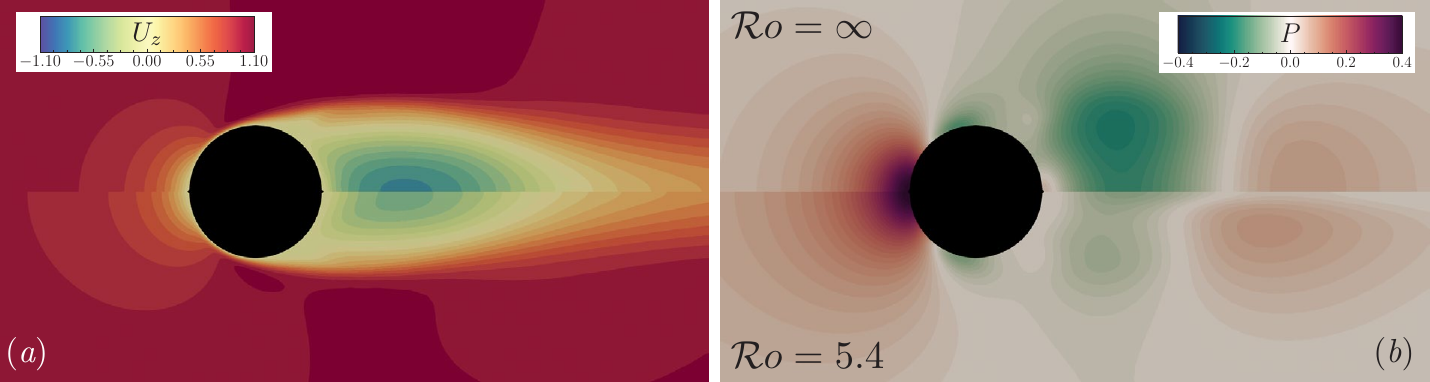}
    \caption{Influence of the rigid-body rotation on the velocity and pressure in the sphere vicinity at high Reynolds number ($\mathcal{R}e=300$). $(a)$: axial velocity (scaled by $U_{\infty}$); $(b)$: pressure (scaled by $\frac{1}{2}\rho U^2_{\infty}$). 
    Upper half in each frame: non-rotating flow ($\mathcal{R}o=\infty$); lower half: $\mathcal{R}o = 5.4$. The  {\color{black} {relative flow with respect to the sphere}} is from left to right.}
    \label{fig:drag_reduc}
\end{figure}

\begin{figure}
    \centering
    \includegraphics[width=0.9\linewidth]{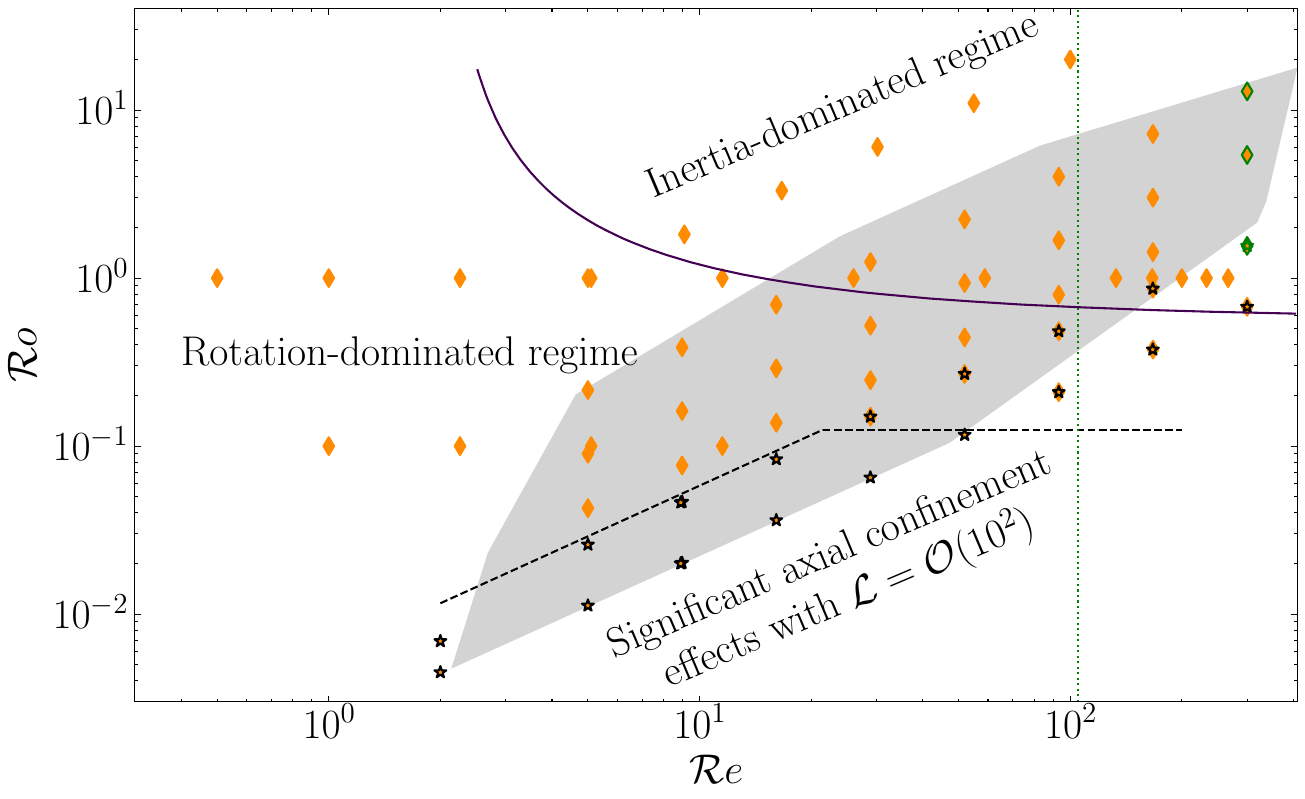}
    \caption{Regimes covered by the present simulations in the parameter space ($\mathcal{R}e, \mathcal{R}o$). \textcolor{orange}{$\blacklozenge$}, $\bigstar$: simulations performed on domains with $\mathcal{L} = 180$ and $\mathcal{L}\approx10^3$ respectively; shaded area: parameter range covered by Maxworthy's 1970 experiments; \textcolor{violet}{------}: $\mathcal{R}o$ vs. $\mathcal{R}e$ as predicted by \eqref{eq:TS} for $C_D\approx6$; 
    - - - -: limit below which confinement effects  are observed with $\mathcal{L}=180$; 
     \textcolor{darkcyan}{$\cdot$ $\cdot\cdot\cdot$}: threshold beyond which the wake is three-dimensional in the limit $\mathcal{R}o\rightarrow\infty$; {\color{black}{lozenges with the green contour correspond to conditions ($\mathcal{R}e=300,\mathcal{R}o>1$) under which the drag is suspected to be significantly affected by three-dimensional effects.}}}
    \label{fig:map}
\end{figure}

In figure \ref{fig:d_vs_re}$(a)$, the drag is observed to be lower than predicted by the standard law at large enough $\mathcal{R}e$ and low enough $\mathcal{T}a$, say $\mathcal{R}e\gtrsim70$ and $\mathcal{T}a\lesssim80$ (consider the last three purple lozenges and the very last dark green lozenge at the bottom right). This is in line with Maxworthy's experimental findings as the bullets confirm, the associated Rossby number  being such that $\mathcal{R}o\gtrsim1$ throughout this regime. The same behaviour was observed numerically by \cite{Rao1995} and \cite{Sahoo2021}. {\color{black}{As non-axisymmetric effects not accounted for in present simulations are known to increase the drag in a non-rotating flow, one might suspect their absence to be at the origin of the low numerical drag values found in the high-$\mathcal{R}e$ range. However axisymmetry in the sphere's wake breaks down only at $\mathcal{R}e\approx105$ when $\mathcal{R}o\rightarrow\infty$ (dashed line in figure \ref{fig:d_vs_re}$(a)$), and the computed drag at $\mathcal{R}e=93$ and $\mathcal{T}a=23.2$ (most left lozenge below the solid line in the figure) is $15\%$ lower than predicted by the standard drag law. Similarly, for the same $\mathcal{T}a$ but $\mathcal{R}e=167$, the drag is $20\%$ smaller than expected on the basis of the standard drag law, whereas the axisymmetric prediction is known to underestimate the drag by only $4\%$ in the limit $\mathcal{R}o\rightarrow\infty$ in that $\mathcal{R}e$-range (see the discussion in \S\,\ref{prelim}). Therefore, one can conclude that the difference observed in the figure is really the result of rotation effects and has the same physical origin as that revealed by Maxworthy's experimental data.}} In the corresponding $\mathcal{R}e$ range, a large standing eddy is present behind the sphere. Figure \ref{fig:drag_reduc} shows how the rigid-body rotation alters the size of this eddy, together with the pressure and axial velocity distributions. Even a modest level of rotation (the lower half of the figure corresponds to $\mathcal{R}o=5.4$)  is seen to reduce significantly the negative axial velocity within the eddy, increasing the pressure in its core. Therefore, compared with the case of a sphere translating in a fluid at rest, the overall pressure difference between the front and rear stagnation points is reduced, lowering the pressure drag. This effect is significant, as the drag may be reduced by $10$ to $20 \%$ with respect to the standard law in the range $100 \lesssim \mathcal{R}e \lesssim 300$. \cite{maxworthy_flow_1970} argued that the pressure increase in the core of the standing eddy is due to the fact that ``\textit{the outward flow of rotating fluid over this} [recirculation] \textit{bubble causes it to rotate at a rate less than the applied value}''. However, present results contradict this explanation. For instance, figure \ref{fig:overview}$(c)$  for $\mathcal{R}e=167,\,\mathcal{T}a=23.2$ shows that the angular velocity within the standing eddy is larger than the applied rotation rate, and the corresponding drag (penultimate purple lozenge in the bottom right corner of figure \ref{fig:d_vs_re}$(a)$) stands $15\%$ below the standard drag curve. Actually, the origin of the drag reduction may be understood by considering the governing equation for the azimuthal vorticity, $\omega_\phi=\partial_zU_\sigma-\partial_\sigma U_z$. 
Rotation enters the $\omega_\phi$-balance through the source term $2\Omega\,\partial_zU_\phi$, similar to that involved in \eqref{vortz}, but with $\partial_zU_z$ replaced with $\partial_zU_\phi$. In the vicinity of the axis, this source term virtually equals $2\Omega\sigma\partial_z\omega_z$. As discussed in \S\,\ref{sec:flow}, $\omega_z$ increases downstream of the sphere with the distance to the rear stagnation point (as figure \ref{fig:overview}$(c)$ confirms). Therefore, this source term is positive within the standing eddy, bringing a positive variation in $\omega_\phi$ compared to the non-rotating configuration. Near the axis, $\omega_\phi\approx-\partial_\sigma U_z$, so that this change in $\omega_\phi$ translates into an increase in $U_z$ as $\sigma\rightarrow0$, i.e. a positive variation of the axial velocity as the rotation axis is approached. Hence, when $U_z$ is negative in the limit $\mathcal{R}o\rightarrow\infty$, finite rotation effects decrease its magnitude, leading to an increase in the local pressure, from which the observed drag reduction ensues.
\\
\begin{figure}
    \centering
    \includegraphics[width = 0.8\linewidth]{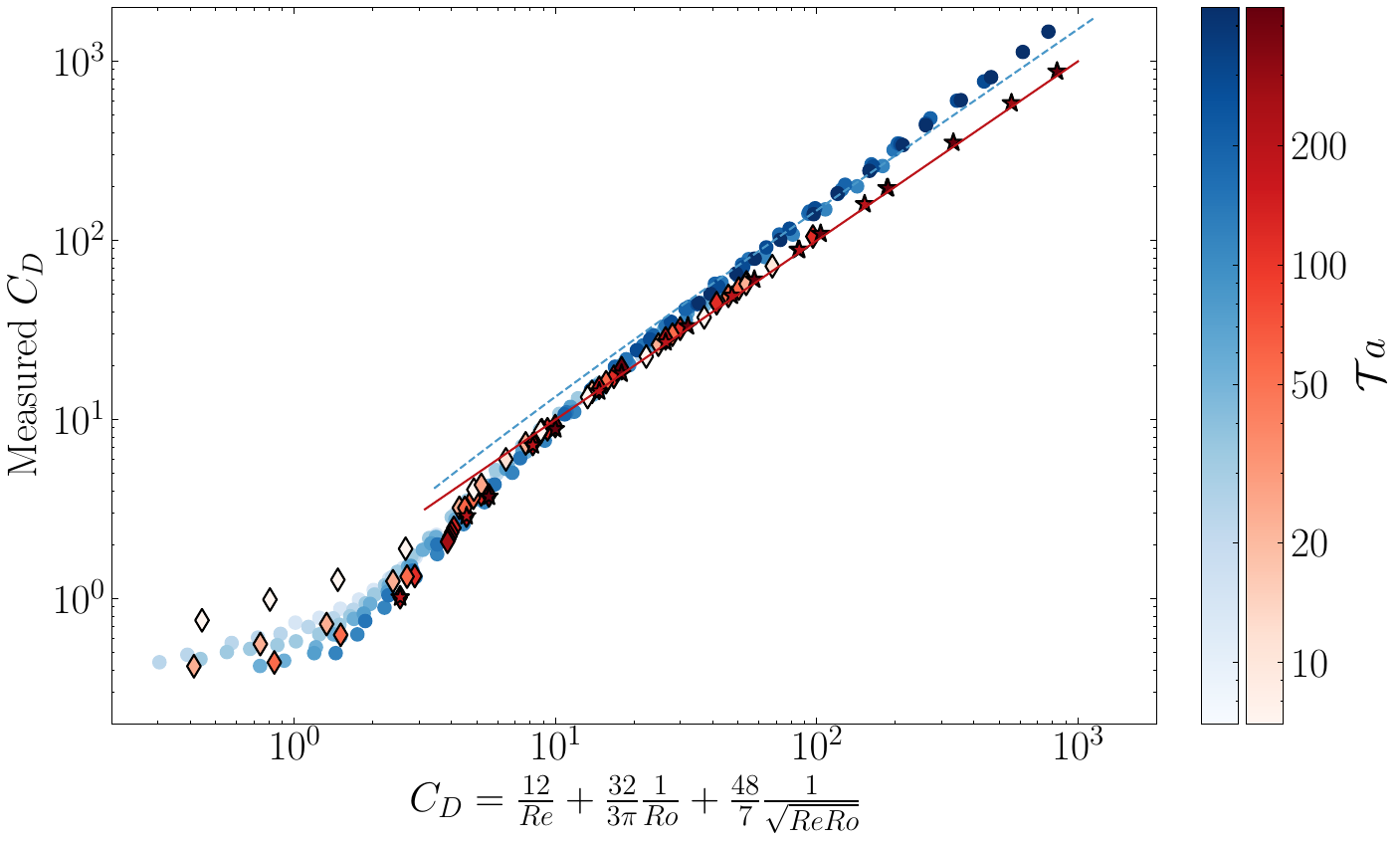}
    \caption{Comparison of the measured drag coefficient with the semi-empirical prediction \eqref{eq:TS}. \textcolor{red}{------}: prediction \eqref{eq:TS}; $\blacklozenge, \bigstar$: present simulations with $\mathcal{L}=180$ and $\approx10^3$, respectively; $\bullet$: experimental data \citep{maxworthy_flow_1970};  \textcolor{black}{- - - -}: `corrected' experimental law \eqref{eq:max_extra}. The red and blue color bars refer to the numerical and experimental data, respectively. In each series, symbols are colored according to the value of $\mathcal{T}a$, darkening as $\mathcal{T}a$ increases.}
    \label{fig:cd_vs_chiste}
\end{figure}
\indent In figure \ref{fig:cd_vs_chiste}, the experimentally and numerically determined drag coefficients are plotted vs the semi-empirical prediction \eqref{eq:TS} suggested by \citet{tanzosh_motion_1994}. This prediction is expected to be valid at arbitrary Taylor number. In contrast, it is only supposed to apply as long as the Reynolds number is low, given the range of validity of \eqref{eq:childress} from which the first two terms of  \eqref{eq:TS} are borrowed. Numerical results are seen to be in excellent agreement with \eqref{eq:TS} as long as $C_D\gtrsim6$, provided the computational domain is long enough. Indeed, following the conclusions of appendix \ref{sec:app_conf}, results corresponding to $C_D\geq10^2$ were obtained using the extended domain with a half-length $\mathcal{L}\approx10^3$, while those corresponding to lower $C_D$ were obtained on the standard domain with $\mathcal{L}=180$. Note that the predictions provided by the two domains match properly throughout the intermediate range {\color{black}{$8\lesssim C_D\lesssim10^2$}}. In stark contrast with present results, Maxworthy's 1970 experimental data stand beyond the prediction \eqref{eq:TS} as soon as $C_D>15$, the difference being up to $50 \%$ for large $C_D$. The empirical extrapolation \eqref{eq:max_extra} supposed to account for axial confinement effects in his device (which had $\mathcal{L}\approx80$ or $120$ depending on the sphere size) only brings a marginal improvement, leaving a $40\%$ over-prediction for $C_D=\mathcal{O}(10^3)$. When $C_D$ is low enough, typically $C_D\lesssim6$, \eqref{eq:TS} is found to overestimate the drag. This regime corresponds to large Reynolds numbers and moderate Rossby numbers. These are the conditions under which the above drag reduction mechanism related to the rotation-induced shortening of the standing eddy operates. The drag modification resulting from this mechanism is obviously not included in the low-$\mathcal{R}e$ asymptotic result \eqref{eq:childress}, making the simple drag law \eqref{eq:TS} inaccurate in this regime. Conversely, \eqref{eq:TS}  is found to hold even in the moderate-to-large Reynolds number regime provided the Rossby number is somewhat lower than unity. For instance, setting $\mathcal{R}o=0.5$ and $\mathcal{R}e=100$ yields $C_D\approx7.9$, which is close to the lower limit of validity of \eqref{eq:TS} according to figure \ref{fig:cd_vs_chiste}. This leads to the conclusion that this simple semi-empirical prediction is actually valid well beyond the low-$\mathcal{R}e$ regime within which it is in principle supposed to hold.\\
\indent That present numerical results closely agree with the semi-empirical prediction \eqref{eq:TS} over two decades of $C_D$ (hence, $\mathcal{R}o$) proves that axial confinement effects are responsible for the long-standing but previously unresolved disagreement between experimental results and theoretical models. The restored agreement obtained by considering stationary axisymmetric solutions of the Navier-Stokes equations also rules out the possibility that the problem could be due to non-axisymmetric or unsteady effects as was previously suggested \citep{minkov_motion_2002}. Although Maxworthy rightly identified the origin of the problem, the correction \eqref{eq:max_extra} he proposed was biased because it was in a good part based on an extrapolation of his previous data obtained in a much shorter device with $5\lesssim\mathcal{L} \lesssim 10.5$, depending on the sphere size \citep{maxworthy_observed_1968}. This extrapolation was not appropriate because end effects in short and long containers do not involve the same mechanisms at all, and therefore do not influence the drag in the same way. In short containers, the direct interaction of the Taylor column with the Ekman layers present along the end walls controls the drag to leading order, making $C_D$ depend on viscosity as \eqref{eq:ms68} shows. This is not the case in long containers, in which the end walls only produce a (mostly inviscid) blocking effect that slightly compresses the Taylor column. This difference induces dramatic consequences on the flow structure in the vicinity of the body. For instance, \citet{ungarish_motion_1995} showed that, for a thin disc, the upstream recirculation exists only if the ratio $\delta=\mathcal{L}/ \mathcal{T}a$ is larger than $0.08$. Maxworthy's 1968 experiments were carried out at very large Taylor numbers, $\mathcal{T}a\geq2.5\times10^3$, so that the corresponding data all correspond to $\delta\leq4\times10^{-3}$, a regime in which the flow within the Taylor column has little to do with that sketched in figure \ref{fig:flow_structure_new} (for which $\delta=0.93$). Because of these structural differences, there was little chance that an extrapolation mixing two fundamentally different regimes could work. 
\\
\indent It is also worth noting that axial confinement effects in Maxworthy's 1970 experiments were actually more severe than can be expected on the basis on the container-to-particle size ratios $\mathcal{L}\approx80$ and $\mathcal{L}\approx120$. Indeed, since the drag was obtained by determining the time of flight of rising particles between two sets of horizontal lines, these particles were closer to the bottom wall when the stopwatch was unlocked and closer to the top wall when it was stopped. Some quantitative details are missing in Maxworthy's description of the experimental protocol. Nevertheless, it may reasonably be hypothesized that the two sets of lines were close to the bottom and upper ends of the `viewing box' that surrounded the {\color{black}{middle}} part of the cylindrical rotating container. With this, it may be estimated that the container length available downstream of the sphere varied over time in the range $48\leq\mathcal{L}\leq112$ for the large particles with which the large-$\mathcal{T}a$ low-$\mathcal{R}o$ conditions were achieved. Obviously, the length available upstream of the particle followed opposite time variations. Therefore, the actual container-to-particle size ratio that determines the strength of confinement effects rather stood in the range $50\lesssim\mathcal{L}\lesssim80$ (grey bars in figures \ref{fig:lus_vs_lx} and \ref{fig:lus_vs_lx_t193}). In contrast, the axial confinement does not vary over time in present computations, since the sphere stays midway between the two end `walls' throughout a run. In appendix \ref{sec:app_conf}, we examine in two low-$\mathcal{R}o$ cases how the drag varies as the length of the computational domain is increased. Based on these variations, we determined the fit \eqref{fitconf} predicting the artificial drag increase induced by axial confinement effects. This fit may be useful to design or interpret future experiments, although some caution is required given the differences between the experimental and numerical setups.\\  
\indent Figure \ref{fig:map} summarizes the various `regimes' encountered in present simulations in the parameter space ($\mathcal{R}e, \mathcal{R}o$), the shaded area sketching the range covered by Maxworthy's 1970 experiments. Three main regions may be identified. Beyond the solid line, inertial effects dominate over those induced by the rigid-body rotation, making the drag depart from the semi-empirical prediction  \eqref{eq:TS}. Below this line, numerical predictions are in good agreement with \eqref{eq:TS}, provided the computational domain is long enough. This constraint is fulfilled with $\mathcal{L}=180$ in between the solid and dashed lines. Confinement effects become more severe below the latter, i.e. for $\mathcal{R}o < 0.125$ when $\mathcal{T}a > 150$, and we had to use the extended domain with $\mathcal{L}\approx10^3$ to get rid of these effects in that range. 
Although figure \ref{fig:cd_vs_chiste} shows that the agreement with \eqref{eq:TS} extends up to the highest Reynolds number considered in the simulations ($\mathcal{R}e=300$) provided $\mathcal{R}o$ is low enough, it must again be stressed that the actual flow is no longer axisymmetric at such Reynolds numbers when rotation effects are moderate or low, as the vertical dotted line in figure  \ref{fig:map}, which corresponds to the transition to three-dimensionality in the limit $\mathcal{R}o\rightarrow\infty$, reminds. {\color{black}{However, as discussed in \S\,\ref{prelim}, three-dimensional effects only marginally affect the drag for $\mathcal{R}e\approx150$ in the non-rotating limit. This is why present results in that range (penultimate vertical series of lozenges in  figure \ref{fig:map}) are still relevant for a comparison with experimental data, and only results corresponding to the three lozenges with the green contour in the rightmost series ($\mathcal{R}e=300$) are expected to be significantly modified by non-axisymmetric effects.}}  

\subsection{Torque}
\label{torque}
As stated in \S\,\ref{sec:equations}, present computations were carried out by imposing that the sphere rotates at the same rate as the undisturbed flow. Therefore, it experiences a nonzero torque and it is of interest to examine how this torque varies with the flow parameters. Making use of the definitions introduced in \S\,\ref{sec:valid}, especially the spherical coordinate system whose origin stands at the sphere centre, the $z$ component of the torque is 
\begin{equation}
M_z={\bf{e}}_z\cdot\int_\mathcal{S}a{\bf{e}}_r\times({\bf{e}}_r\cdot{\bf{T}}\big|_{r=a})dS\,,
\label{torque0}
\end{equation}
where $\mathcal{S}$ denotes the sphere surface. Expanding the surface traction ${\bf{e}}_r\cdot{\bf{T}}\big|_{r=a}$ component-wise, \eqref{torque0} is found to reduce to
\begin{equation}
M_z=-\rho\nu a\int_\mathcal{S}({\bf{e}}_z\cdot{\bf{e}}_\theta)\partial_rU_\phi dS=2\pi\rho\nu a^3\int_0^\pi\sin^2\theta(\partial_rU_\phi)\big|_{r=a} d\theta\,.
\label{torque1}
\end{equation}
Noting that $\sin^2\theta$ is symmetric with respect to the equatorial plane $\theta=\pi/2$, it is relevant to expand $U_\phi$ into a component that shares this property, i.e. an even function of $z$, and a component that is antisymmetric with respect to the equatorial plane, i.e. an odd function of $z$ (formally, this could be achieved \textit{via} Fourier transform for instance). Only the even component of $U_\phi$ contributes to $M_z$. As such a component results from the downstream advection of the negative upstream axial vorticity (see frames $(a)-(b)$, $(e)-(f)$ and $(h)-(i)$ in figure \ref{fig:overview_iso}), $M_z$ is expected to be negative, which in the case of a torque-free sphere would make it rotate slower than the undisturbed fluid. Thus, it is appropriate to introduce a torque coefficient, $C_T$, related to the axial torque through $M_z=-\frac{\pi}{2}C_Ta^3\rho U_\infty^2$. Based on this definition and on the above remark, one has
\begin{equation}
C_T=-4\mathcal{R}e^{-1}\int_0^\pi\sin^2\theta(\partial_{r/a}U_\phi^e)\big|_{r=a} d\theta\,,
\label{torque2}
\end{equation}
where $U_\phi^e$ stands for the dimensionless even contribution to $U_\phi$.\\
\indent 
Outside the boundary layer, the dimensionless thickness of which is denoted as $\delta_B$, one has $U_\phi^e\approx(\sigma/a)\omega_z^e$, where $\omega_z^e$ stands for the (dimensionless) even component of $\omega_z$. Since $U_\phi^e=0$ at the sphere surface and $\sigma/a\approx 1$ in the equatorial region, $(\partial_{r/a}U_\phi^e)\big|_{r=a}\sim\omega_z^e/\delta_B$, so that $C_T\sim\mathcal{R}e^{-1}\omega_z^e/\delta_B$. To determine the scaling laws obeyed by $C_T$, one must consider the governing equation \eqref{vortz} for $\omega_z$, keeping in mind that $\delta_B\sim\mathcal{T}a^{-1/2}$ if $\mathcal{R}o\ll1$ and $\mathcal{T}a\gg1$, while $\delta_B\sim\mathcal{R}e^{-1/2}$ in the opposite limit $\mathcal{R}o\gg1$, $\mathcal{R}e\gg1$. In the latter regime, making use of the near-axis approximation discussed in \S\,\ref{sec:flow} for the tilting term $-\omega_\sigma\partial_\sigma U_z$, the $\omega_z$-balance outside the boundary layer reduces at leading order to $U_\sigma\partial_\sigma\omega_z +(U_z+\sigma\partial_\sigma U_z)\partial_z\omega_z-\omega_z\partial_zU_z\approx2\Omega\partial_zU_z$. Since the axial velocity at radial positions $\sigma/a\approx 1$ decreases (resp. increases) as $z$ increases upstream (resp. downstream) of the sphere, the leading-order contribution to $\partial_zU_z$ is an odd function of $z$. Hence, the flow past the sphere is dominated by an even component in $U_z$ and, owing to continuity, an odd component in $U_\sigma$. Then, according to the above form of the $\omega_z$-balance, it turns out that the leading contribution to $\omega_z$ is even with respect to $z$. Effects of the Coriolis force do not put a severe restriction on the variations of the flow field in the $z$ direction in that regime. Therefore, outside the boundary layer, $\partial_z \sim a^{-1}\sim\partial_\sigma, U_z\sim U_\infty, U_\sigma\sim a\delta_BU_\infty$ and the $\omega_z$-balance implies $U_\infty\omega_z/a\sim\Omega U_\infty/a$, i.e. $\omega_z\sim\Omega$, so that $\omega_z^e\sim\mathcal{R}o^{-1}$. Hence $\omega_z^e/\delta_B\sim \mathcal{R}e^{1/2}\mathcal{R}o^{-1}$ and 
\begin{equation}
C_T\big|_{\mathcal{R}o\gg1,\,\mathcal{R}e\gg1}\sim\mathcal{R}o^{-1}\mathcal{R}e^{-1/2}\,.
\label{CT1}
\end{equation}
 Let us now consider the low-$\mathcal{R}o$ limit in which significant axial variations of the flow field exist only within the Ekman layer. The dominant balance for $\omega_z$ then reads $2\Omega\partial_zU_z\approx-\nu\nabla^2\omega_z$. Near the equatorial plane, axial variations at radial positions $\sigma\approx a$ in the Ekman layer take place over distances of the order of the sphere radius, so that $\partial_z\approx a^{-1}$. The axial velocity being of the order of $U_\infty$ at the outer edge of that layer, one has $\partial_zU_z\sim U_\infty/a$. 
 Since $U_z$ is almost symmetric with respect to the sphere's equator, the dominant contribution to the source term in the $\omega_z$-balance is an odd function of $z$, and so is the leading contribution to $\omega_z$, say $\omega_z^oU_\infty/a$. The scaling of $\omega_z^o$ results from the balance $2\Omega U_\infty/a\sim\nu (U_\infty/a)\omega_z^o/(a\delta_B)^2$, which yields $\omega_z^o\sim1$. If the Rossby number is small but finite, advection past the sphere brings a small correction to the $\omega_z$-distribution through the term  $\{U_\sigma\partial_\sigma\omega_z^o +(U_z+\sigma\partial_\sigma U_z)\partial_z\omega_z^o-\omega_z^o\partial_zU_z\}U_\infty/a$, which is almost an even function of $z$. To balance this term, an even correction to $\omega_z$ is required, say $\omega_z^eU_\infty/a$, and is provided by the corresponding viscous term, $\nu( U_\infty/a)\nabla_{\sigma,z}^2\omega_z^e$. Still in the vicinity of the equatorial plane, $\partial_\sigma\sim(a\delta_B)^{-1}$ in the Ekman layer, and $U_\sigma\sim\delta_B U_\infty$ at its outer edge. Therefore, the above inertial source term is dominated by the contribution $U_\infty(\sigma/a)\partial_\sigma U_z\partial_z\omega_z^o\sim (U_\infty/a)U_\infty/(a\delta_B)\omega_z^o$ and the above balance implies $U_\infty/(a\delta_B)\omega_z^o\sim\nu\omega_z^e/(a\delta_B)^2$, i.e. $\omega_z^e\sim\mathcal{R}e\delta_B\,\omega_z^o$. According to the scalings obeyed by $\delta_B$ and $\omega_z^o$, this yields  $\omega_z^e\sim\mathcal{R}e\mathcal{T}a^{-1/2}$. Hence $\omega_z^e/\delta_B\sim \mathcal{R}e$ and 
 \begin{equation}
C_T\big|_{\mathcal{R}o\ll1,\,\mathcal{T}a\gg1}\sim1\,,
\label{CT2}
\end{equation}
indicating that the torque coefficient is now independent of the control parameters. Therefore, \eqref{CT1} and \eqref{CT2} predict that $C_T$ exhibits two different scaling laws, according to the magnitude of the Rossby and Reynolds numbers. In rotation-dominated regimes, where advective effects only provide a small correction to the dominant axial vorticity balance, $C_T$ is constant, whereas it decays with both $\mathcal{R}o$ and $\mathcal{R}e$ in advection-dominated regimes.

\begin{figure}
    \centering
    \includegraphics[width = 0.8 \textwidth]{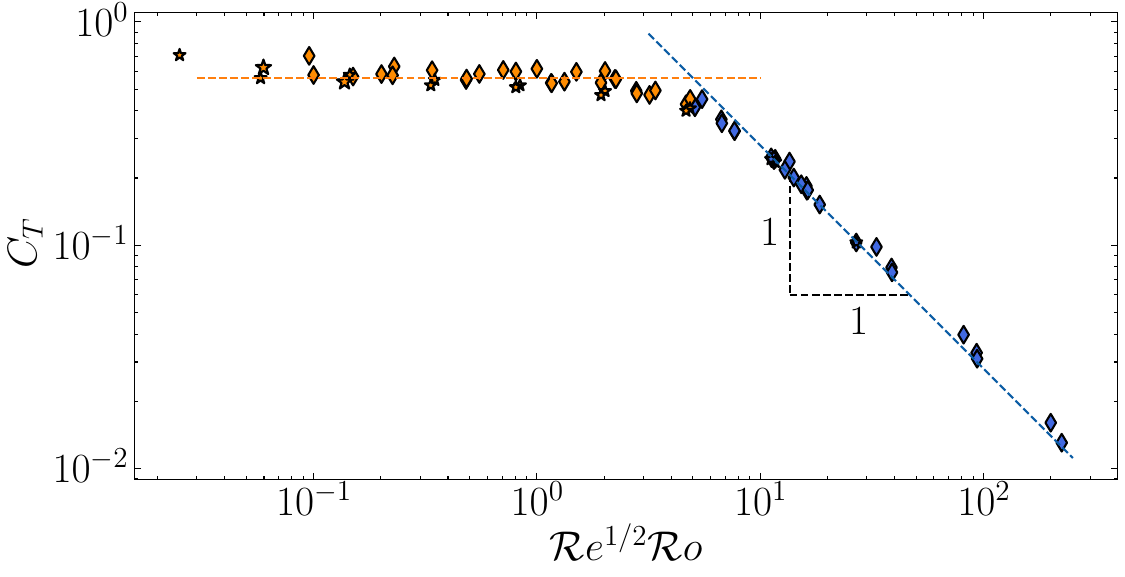}
    \caption{Torque coefficient $C_{T}$ as a function of $\mathcal{R}o\mathcal{R}e^{1/2}$. $\blacklozenge$ and $\bigstar$: results obtained  on domains with $\mathcal{L}=180$ and $\mathcal{L}\approx10^3$, respectively. Blue (resp. orange) colours: configurations corresponding to $C_D<6$ (resp. $C_D\geq6$).
    }
    \label{fig:torque}
\end{figure}


Figure \ref{fig:torque} shows how numerical results compare with the above predictions. 
As long as $\mathcal{R}o\mathcal{R}e^{1/2}$ is less than $\approx5$, the torque coefficient agrees with the prediction \eqref{CT2}, with $C_T= 0.56 \pm 0.06$. Beyond $\mathcal{R}o\mathcal{R}e^{1/2}\approx5$, i.e. when inertial effects are moderate to large and effects of rotation are moderate or weak, $C_T$ is closely approximated by the decay law $C_T=2.8\,\mathcal{R}o^{-1}\mathcal{R}e^{-1/2}$, in agreement with \eqref{CT1}. 
As the colours of the symbols in figure \ref{fig:torque} indicate, the first regime coincides with that in which figure \ref{fig:cd_vs_chiste} revealed that the drag coefficient agrees well with the prediction \eqref{eq:TS}. Conversely, the second regime is that in which $C_D$ departs from this prediction. \\
\indent The numerical results for the torque coefficient may be used to estimate the differential rotation of the sphere, say $\Omega_s$, required to satisfy the torque-free condition. Indeed, this rotation induces an azimuthal velocity $\Omega_sa\sin\theta$ at the sphere surface, so that the dimensionless velocity gradient involved in the definition \eqref{torque2} of $C_T$ becomes $(\partial_{r/a}U_\phi^e)\big|_{r=a}\sim(\omega_z^e-\frac{a\Omega_s}{U_\infty}\sin\theta)/\delta_B$. The approximate change in the torque coefficient is then $4(\mathcal{R}e\delta_B)^{-1}\frac{a\Omega_s}{U_\infty}\int_0^\pi\sin^3\theta d\theta=\frac{16}{3}(\mathcal{R}e\delta_B)^{-1}\frac{a\Omega_s}{U_\infty}$. Inspection of figure \ref{fig:slices}$(i)$ allows the distance to the sphere surface at which the swirl velocity $U_\phi/\sigma$ reaches its extremum in the equatorial plane to be determined. This leads to the approximate estimates $\delta_B\approx2.0\,\mathcal{T}a^{-1/2}$ and $\delta_B\approx2.0\,\mathcal{R}e^{-1/2}$ in the low- and high-$\mathcal{R}o$ regimes, respectively. In the former regime, numerical results showed that $C_T\approx0.56$, so that the torque-free condition is achieved with $\frac{a\Omega_s}{U_\infty}\approx-\frac{3}{16}\times0.56\,\mathcal{R}e\,\delta_B\approx-0.2\,\mathcal{R}e\mathcal{T}a^{-1/2}$. Similarly, in the high-$\mathcal{R}o$ regime, we found $C_T\approx2.8\,\mathcal{R}o^{-1}\mathcal{R}e^{-1/2}$, so that $\frac{a\Omega_s}{U_\infty}\approx-1.0\,\mathcal{R}o^{-1}$. Normalized with respect to the imposed rigid-body rotation rate, these estimates become 
\begin{equation}
\frac{\Omega_s}{\Omega}\big|_{\mathcal{R}o\ll1,\,\mathcal{T}a\gg1}\approx-0.2\,\mathcal{R}o^{3/2}\mathcal{R}e^{1/2}\quad\mbox{and }\quad \frac{\Omega_s}{\Omega}\big|_{\mathcal{R}o\gg1,\,\mathcal{R}e\gg1}\approx-1.0\,. 
\label{rotdif}
\end{equation}
The differential rotation is predicted to be very small in the low-$\mathcal{R}o$ regime, with for instance $\Omega_s/\Omega=-1.7\times10^{-3}$ in the configuration of figure \ref{fig:overview_iso}$(g)$. In contrast, the sphere is predicted to have a negligible rotation with respect to the laboratory frame when the Rossby and Reynolds numbers are both large (frames $(b)$, $(c)$ and $(f)$ in figure \ref{fig:overview_iso}). Of course, these are only rough estimates, since we used crude approximations to evaluate the velocity gradient in \eqref{torque2}, and the sphere rotation is expected to induce slight modifications in the distribution of the azimuthal vorticity in the sphere vicinity. Let us also mention that, in the limit $\mathcal{R}e\ll\mathcal{T}a^{1/2}\ll1$, \citet{childress_slow_1964} predicted $\Omega_s/\Omega\approx-\frac{1}{5}\mathcal{R}e^{2}\mathcal{T}a^{-1/2}$. This prediction differs from those obtained here, in particular in the low-$\mathcal{R}o$ regime, since the first estimate in \eqref{rotdif} may be rewritten in the form $\Omega_s/\Omega\approx-0.2\,\mathcal{R}e^{2}\mathcal{T}a^{-3/2}$, which corresponds to the same dependence with respect to $\mathcal{R}e$ but a faster decay with $\mathcal{T}a$. This is no surprise, as we assume the Taylor number to be large, which makes all processes governing the body rotation controlled by the Ekman layer when $\mathcal{R}o$ is low. In contrast, Childress' analysis assumes $\mathcal{R}e\ll\mathcal{T}a^{1/2}\ll1$, so that inertial effects responsible for this differential rotation manifest themselves only at large $\mathcal{O}(\mathcal{T}a^{-1/2})$ dimensionless distances from the body. \\
\begin{table}
	\centering
\begin{tabular}{clc l cl l  l clc l l clc l l clc l l clc l l clc l l c l c l }
	\hline
	Case     & $\mathcal{T}a$& $\mathcal{R}o$ & $\mathcal{R}o\mathcal{R}e^{1/2}$&$C_D$ &$\Delta C_D$&$C_T$ &$\Omega_s/\Omega$\\
 \hline
	\multirow{2}{*}{$(g)$}         &\hfil \multirow{2}{*}{$445$}  & \hfil \multirow{2}{*}{0.02} & \hfil\multirow{2}{*}{ 0.06} & 196.2 &\hfil \multirow{2}{*}{ 0.05}& \hfil0.63 &\hfil--  \\
	            &  &   & & 196.1 &&\hfil-- & \hfil-0.0015      \\
	\multirow{2}{*}{$(d)$}         &\hfil \multirow{2}{*}{$117$} &\hfil  \multirow{2}{*}{0.076} &\hfil \multirow{2}{*}{0.227} & 58.42 &\hfil\multirow{2}{*}{0.15} &\hfil 0.632 &\hfil --   \\
	            &  &   & & 58.33 &&\hfil-- & \hfil-0.0098      \\
       \multirow{2}{*}{$(h)$}         &\hfil \multirow{2}{*}{$445$}  &\hfil  \multirow{2}{*}{0.117} &\hfil \multirow{2}{*}{0.843} & 33.32 &\hfil\multirow{2}{*}{0.36} &\hfil 0.52 &\hfil --   \\
	            &  &   & & 33.20 &&\hfil-- &\hfil -0.042      \\
\multirow{2}{*}{$(a)$}         & \hfil\multirow{2}{*}{$23.2$} & \hfil \multirow{2}{*}{0.384} &\hfil \multirow{2}{*}{1.145} & 14.27 &\hfil\multirow{2}{*}{0.92} &\hfil 0.53 &\hfil --   \\
	            &  &   & & 14.14 &&\hfil-- &\hfil -0.067      \\
\multirow{2}{*}{$(e)$}         & \hfil\multirow{2}{*}{$117$} & \hfil \multirow{2}{*}{0.444} &\hfil \multirow{2}{*}{ 3.20 } & 8.84 &\hfil \multirow{2}{*}{1.38} &\hfil 0.471 &\hfil --   \\
	            &  &   & & 8.72 &&\hfil-- &\hfil -0.26      \\
  \multirow{2}{*}{$(i)$}         & \hfil\multirow{2}{*}{$445$} &\hfil  \multirow{2}{*}{0.375} &\hfil \multirow{2}{*}{4.846} & 8.91 &\hfil \multirow{2}{*}{-1.44} &\hfil 0.411 &\hfil --   \\
	            &  &   & & 9.04 &&\hfil-- &\hfil -0.37      \\
\multirow{2}{*}{$(b)$}         &\hfil \multirow{2}{*}{$23.2$} &\hfil  \multirow{2}{*}{2.24} & \hfil\multirow{2}{*}{16.15} & 1.25 &\hfil \multirow{2}{*}{7.89}&\hfil 0.18 &\hfil --   \\
	            &  &   & & 1.14 &&\hfil-- &\hfil -0.72      \\
  \multirow{2}{*}{$(f)$}         &\hfil \multirow{2}{*}{$117$} &\hfil  \multirow{2}{*}{1.43} &\hfil \multirow{2}{*}{18.48} & 1.34 &\hfil \multirow{2}{*}{9.84} &\hfil 0.152 &\hfil --   \\
	            &  &   & & 1.22 &&\hfil-- &\hfil -0.81      \\
\multirow{2}{*}{$(c)$}         & \hfil\multirow{2}{*}{$23.2$} &\hfil  \multirow{2}{*}{7.2} &\hfil \multirow{2}{*}{93.05} & 0.558 &\hfil\multirow{2}{*}{2.76} &\hfil 0.033 &\hfil --   \\
	            &  &   & & 0.543 &&\hfil-- &\hfil -0.86      \\
\hline
\end{tabular}
\caption{Influence of the sphere rotation on the drag. Results are sorted by increasing values of $\mathcal{R}o\mathcal{R}e^{1/2}$. The labels in the first column refer to the frames in figure \ref{fig:overview_iso}; for each set of conditions, the drag coefficient in the first row  ($C_D^{0}$) was obtained with $\Omega_s=0$, while that in the second row ($C_D^{tf}$) corresponds to the torque-free condition; $\Delta C_D=(C_D^0/C_D^{tf}-1)\times100$ is the percent difference between the two drag coefficients. Cases $(g)-(i)$ were computed using the extended domain with $\mathcal{L}\approx10^3$.}
\label{torquefree}
\end{table}
\indent To check the above prediction for $\Omega_s$ and assess the influence of this rotation on the drag, we carried out additional simulations corresponding to the torque-free condition. This condition was enforced iteratively, and convergence was considered to be reached when the final torque was less than $1\%$ of its stationary value in the case $\Omega_s=0$. We ran these simulations for the nine cases for which the flow structure is displayed in figure \ref{fig:overview_iso}. These configurations span the range of conditions considered in this work, especially the two regimes  exhibited in figure \ref{fig:torque}. Results of these simulations are summarized in table \ref{torquefree}. It is seen that the sphere rotation has a negligible influence on the drag (i.e. the two values of $C_D$ differ by less than $2\%$) in all cases with $\mathcal{R}o\lesssim0.5$. This influence is larger with moderate to low rotation levels, as could be expected on the basis of the $\mathcal{O}(1)$ values of $\Omega_s/\Omega$ predicted by \eqref{rotdif} in this regime. Nevertheless the relative difference between the two $C_D$ never exceeds $10\%$. Hence, replacing results of figures \ref{fig:d_vs_re} and \ref{fig:cd_vs_chiste} obtained with $\Omega_s=0$ with those corresponding to the torque-free condition makes no visual difference, even in the most inertial regimes. The estimates \eqref{rotdif} predict $\Omega_s/\Omega=-0.0017$ in case $(g)$ and $\Omega_s/\Omega=-1.0$ in case $(c)$, which compares well with the values reported in the top and bottom lines of table \ref{torquefree}. 



\section{Summary and concluding remarks} \label{sec:concl}

With the aid of numerical simulations, we revisited the classical problem of a rigid sphere steadily translating along the axis of a rotating container filled with a slightly viscous fluid. Assuming the flow to be axisymmetric and the sphere to rotate at the same rate as the container, we considered a large number of combinations in the range $\mathcal{T}a \in [20, 450]$ and $\mathcal{R}e \in [5, 300]$, covering the Rossby number range $\mathcal{R}o \in [10^{-2}, 10]$. These conditions correspond to those explored experimentally by Maxworthy in his 1970 reference study \citep{maxworthy_flow_1970}.\\
\indent Although the problem looks easy from a numerical point of view by today's standards, it is actually challenging regarding the computational domain and the discretization grid. The reason is that the flow has to be captured accurately both in the thin Ekman boundary layer surrounding the body and over very long distances upstream and downstream of it in the near-axis region corresponding to the Taylor column. This is presumably the reason why it took half a century to repeat Maxworthy's experiments on a computer. We dealt with this technical issue by making use of a boundary-fitted orthogonal curvilinear grid that combines the advantages of spherical coordinates in the sphere vicinity with those of cylindrical coordinates far from it. \\
\indent Thanks to the design of this grid, the characteristics of the flow could be examined in detail throughout the desired parameter range,  and several quantities were compared quantitatively with available predictions. In particular, we could observe the inertial wave pattern radiated by the sphere, and check that the associated wavelength agrees well with the inviscid theoretical prediction $\Lambda=\pi\mathcal{R}o$ \citep{taylor_1922}. We also examined how the characteristics of the flow within the Taylor column vary with the control parameters. In particular, we found that, for $\mathcal{T}a\gtrsim100$, the length of the upstream recirculation region follows the law $\ell_s=0.052\,\mathcal{T}a$ established by \cite{tanzosh_motion_1994} in the zero-$\mathcal{R}o$ limit. Using horizontal slices of the three velocity components at various altitudes, we could also clarify some interesting low-$\mathcal{R}o$ mechanisms, such as that leading gradually to a plug-like distribution of the angular swirl as one moves axially away from the body through the \textcolor{black}{nearly} geostrophic and recirculation regions. Slices in the equatorial plane also helped to highlight some consequences of inertial effects that break the symmetries inherent to the zero-Rossby-number limit. While these symmetries impose that the radial and azimuthal velocities are zero in that plane at $\mathcal{R}o=0$, we found that these components develop large negative peaks within the Ekman layer, with respective minima of the order of $20\%$ and $60\%$ of the sphere speed for $\mathcal{R}o\approx0.5$. Conversely, these effects drastically reduce the magnitude of the large positive peak of the axial velocity encountered in that layer in the zero-$\mathcal{R}o=0$ limit, dividing it by a factor of two for $\mathcal{R}o\approx0.5$, which of course has direct consequences on the fluid exchange between the fore and aft Taylor columns. \\
\indent We determined the drag experienced by the sphere for a large number of $(\mathcal{T}a, \mathcal{R}e)$ sets and performed a systematic comparison of the numerical results with Maxworthy's 1970 data and available predictions. Comparing low-$\mathcal{R}o$ large-$\mathcal{T}a$ results obtained on computational domains having $\mathcal{L}=\mathcal{O}(10^2)$ with the zero-$\mathcal{R}o$ predictions of \cite{tanzosh_motion_1994} based on a boundary-integral approach (hence, an infinite domain) made it clear that axial confinements effects dramatically enhance the drag in this regime, owing to the slight changes they induce in the structure of the Taylor column. To get rid of almost all of this undesired influence, we designed a grid with $\mathcal{L}=\mathcal{O}(10^3)$ on which the drag coefficients were found to agree with the zero-$\mathcal{R}o$ prediction within a few percent. Hence, this extreme sensitivity of the drag to axial confinement effects is the reason why Maxworthy's 1970 data (obtained in a container with $\mathcal{L}\approx80$) stand systematically and significantly beyond theoretical predictions. 
Once these effects are eliminated, the drag coefficient agrees well with the semi-empirical law \eqref{eq:TS} that accounts for the combined effects of rotation, viscosity and weak inertia. Actually, the domain of validity of \eqref{eq:TS} was found to extend throughout the range of conditions under which $C_D\gtrsim 6$. Hence, we could conclude that \eqref{eq:TS} is valid up to $\mathcal{R}e= \mathcal{O}(10^2)$, provided rotation effects are large enough. In contrast, \eqref{eq:TS}  overestimates the drag when inertial effects effects are `too' dominant. Remarkably, in this high-$\mathcal{R}e$ and moderate-to-large $\mathcal{R}o$ regime, the drag is also overestimated by the standard law designed for a sphere translating in a fluid at rest. The reason for this could be ascribed to the influence of (weak) rotation effects on the azimuthal vorticity in the sphere wake. Rotation contributing to increase this vorticity component in that region, it weakens the negative axial fluid velocity within the standing eddy, and therefore reduces the pressure drag, a scenario confirmed by the numerical velocity and pressure distributions at the back of the sphere. \\
\indent Since the sphere was assumed to rotate at the same rate as the undisturbed fluid, we could determine the torque it experiences. It turned out that this torque obeys two different scaling laws, depending on the flow regime. The torque coefficient is constant when rotation effects are dominant, more precisely as long as $\mathcal{R}o\mathcal{R}e^{1/2}\lesssim5$. In contrast, this coefficient decays as $\mathcal{R}o^{-1}\mathcal{R}e^{-1/2}$ in inertia-dominated regimes. Interestingly, the conditions corresponding to the transition between the two scalings coincide with the threshold below which the drag law \eqref{eq:TS} ceases to be valid, i.e. $C_D\approx6$. The two scaling laws were rationalized by examining the axial vorticity balance in the sphere vicinity and the associated symmetries with respect to the equatorial plane, from which the dominant scalings governing the symmetric vorticity component, which originates in advective effects, could be determined. Numerical results for the torque were used to infer the differential rotation of the sphere achieving the torque-free condition. It was found that the differential rotation rate, normalized by the rotation rate of the outer fluid, scales as $\mathcal{R}o^{3/2}\mathcal{R}e^{1/2}$  in the rotation-dominated regime, while it becomes constant in inertia-dominated regimes. Some simulations were carried out under the torque-free condition. They revealed virtually no influence of the sphere rotation on the drag as long as the Rossby number is less than unity, and a modest influence, with relatives differences $\lesssim10\%$, in the most inertial regimes. \vspace{2mm}\\
\indent The present work calls for several extensions in at least three directions. First, three-dimensional effects were deliberately ignored here. Although their influence on the drag is presumably marginal in the parameter range we explored, this secondary effect is worth quantifying. From a more fundamental point of view, determining the threshold $\mathcal{R}e_c(\mathcal{R}o)$ beyond which the wake becomes three-dimensional, the nature of the corresponding bifurcation and the spatial structure of the first  three-dimensional mode would be a significant addition to the current knowledge concerning high-$\mathcal{R}e$ low-to-moderate-$\mathcal{R}o$ flows past axisymmetric bodies. Numerical tools designed to perform global linear stability analysis in axisymmetric open flows are now mature and could be easily adapted to tackle this problem. \\
\indent Second, although only results concerning the steady-state configuration were reported here, transient regimes are also worthy of investigation. In particular, examining how the flow structure and the drag change when the rotation rate is suddenly increased or decreased at a given Reynolds number is a relevant question to predict transient effects in rapidly rotating suspensions and centrifugation processes. Since such a variation induces a change in the drag, the settling or rise speed of the particle also varies. The force balance governing the velocity of particles moving under time-dependent conditions in a viscous fluid is usually split into several distinct contributions, although this splitting is only rigorously justified under creeping-flow conditions. Besides the net body weight and the steady (or quasi-steady) drag, one then finds an added-mass force that opposes the relative acceleration between the particle and fluid, and a history force resulting from the unsteady transport of vorticity past the particle. The added-mass effect being due to the no-penetration of the fluid across the body surface, the corresponding force depends only on the instantaneous relative acceleration and on the body shape. Hence, for a given acceleration, it is unaffected by rotation effects, a conclusion that we could confirm numerically \citep{Auregan2020}. Things are different regarding the history contribution, the evolution of which depends on the past history of the relative acceleration weighted by a time-dependent kernel. This kernel expresses the way a change in the vorticity at the particle surface propagates in the flow under the combined effect of viscosity, inertia, and possible nonconservative forces, here the Coriolis force. As such, this kernel is expected to depend on the Rossby number. In the aforementioned preliminary investigation, we could verify that this is indeed the case. Therefore, a systematic study of history effects in the presence of rigid-body rotation appears to be an important objective for future work. Such an investigation should presumably combine a theoretical approach in the zero-$\mathcal{R}o$ limit with numerical simulations to explore the influence of finite advective effects. \\
\indent Last but not least, drops and bubbles offer challenging additional questions. The theoretical investigations of \cite{Bush1992} (in short containers) and \cite*{Bush1995} (in both short and long containers) performed in the zero-$\mathcal{R}o$ limit provide interesting insights into the effects of the drop-to-fluid viscosity ratio and the centrifugal-to-surface tension force ratio (so-called rotational Bond number). The drops were shown to take prolate shapes due to the centrifugal force;  the larger the rotational Bond number, the more the drop elongates along the rotation axis. Remarkably, in long containers, the rise or settling speed was predicted to be nearly independent of the drop viscosity and detailed shape, and to depend essentially on its equatorial radius. The reason is that the drag directly results from the efficiency with which the fluid is transported from the fore to the aft Taylor column and, in long containers, this transport mostly takes place through the Stewartson layer rather than \textit{via} the Ekman boundary layer. How these features are modified by advective effects is currently essentially unknown\textcolor{black}{, but these effects are suspected to be in good part responsible for the significant overestimate of the drag predicted in the zero-$\mathcal{R}o$ approximation, compared to experimental data. \cite{ungarish1996some} introduced a `quasi-geostrophic' approximation incorporating some finite inertial corrections to remedy this problem, but this  refinement only slightly reduced the disagreement.} These are some of the reasons why the investigation carried out here should be repeated with drops and bubbles. Although this is technically challenging, a variety of numerical approaches now allow the efficient and accurate treatment of boundary conditions at a deformable interface with finite surface tension. Hence, exploring how the zero-$\mathcal{R}o$ findings are altered by the presence of finite advective effects appears as an exciting and reachable continuation of the present work. \\


{\textbf {Acknowledgment}} 

The authors thank Prof. Marius Ungarish for stimulating discussions that contributed to motivate the present study, and for useful comments on the original version of the manuscript. Part of this work was performed using HPC resources from CALMIP (Grant 2020-[P1525]).  \\

{\textbf {Declaration of interests.}} The authors report no conflict of interest.\\

{\textbf {Author ORCIDs.}}\\
T. Aur\'egan https://orcid.org/0000-0001-6301-9006;\\
T. Bonometti https://orcid.org/0000-0001-6869-553X;\\
J. Magnaudet https://orcid.org/0000-0002-6166-4877.

\appendix
\section{Grid design} \label{sec:conv_mesh}
\begin{figure}
    \centering
    \includegraphics[width=0.5\textwidth]{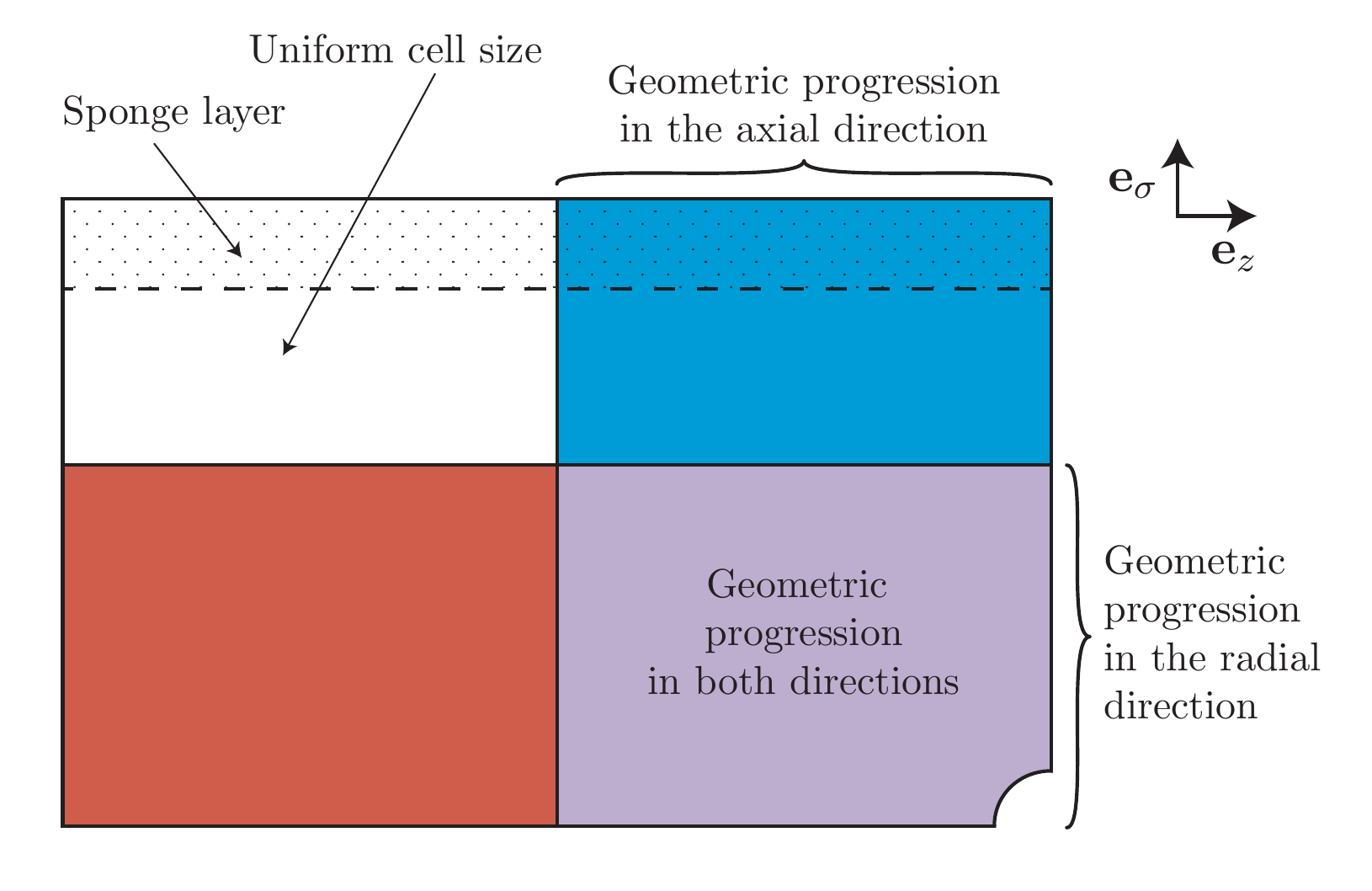}
    \caption{Sketch of the different zones of the grid. }
    \label{fig:zones}
\end{figure}

\begin{figure}
    \centering
    \includegraphics[width=0.8\textwidth]{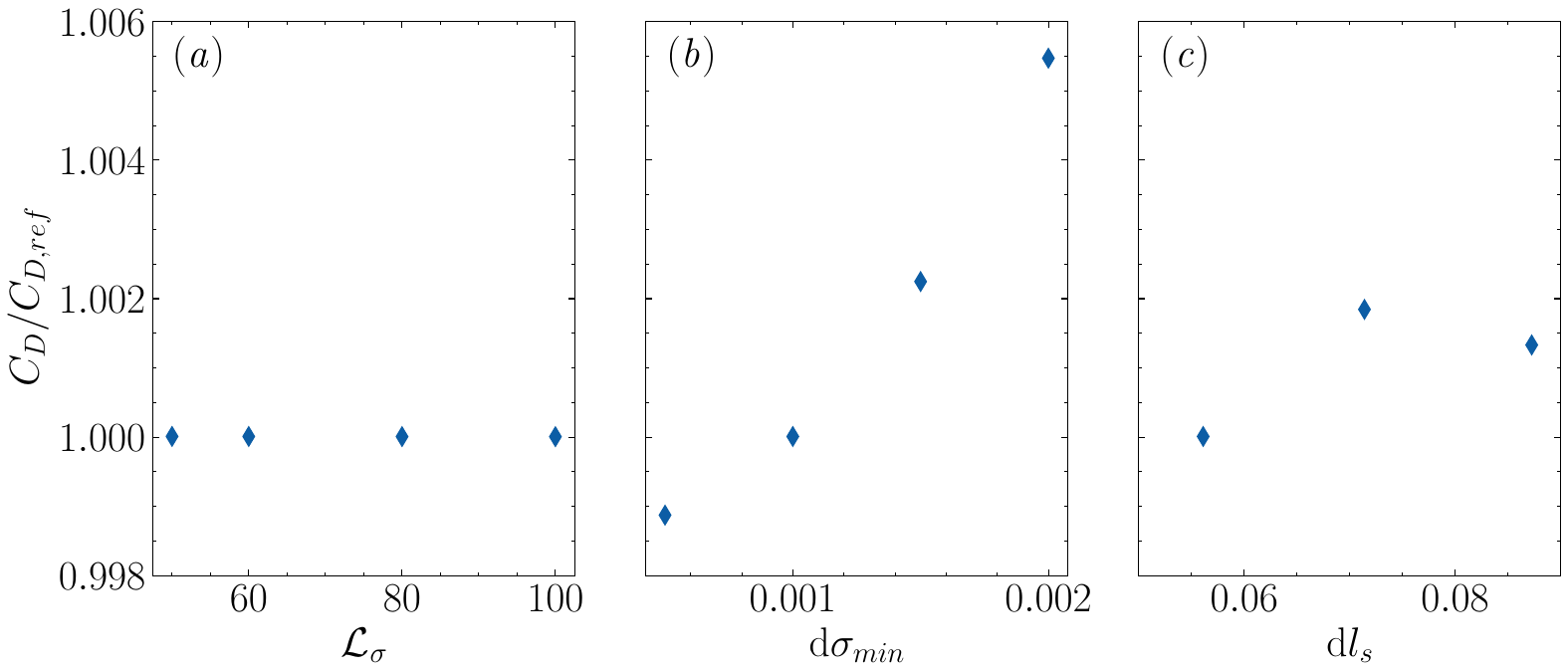}
    \caption{Sensitivity of the drag to some characteristics of the grid for $\mathcal{R}o = 0.1$ and $\mathcal{T}a = 400$. $(a)$: influence of the radius $\mathcal{L}_\sigma$ of the computational domain; $(b)$: influence of the thickness of the first row of cells along the sphere surface and the rotation axis ($\mathrm{d}\sigma_{min}$ is the corresponding thickness at the downstream and upstream ends of the non-uniform region); $(c)$: influence of the length $\mathrm{d}l_{s}$ of the cells adjacent to the sphere surface. The reference drag coefficient, $C_{D, ref}$, is obtained with the standard values $\mathcal{L}_\sigma=60$, $\mathrm{d}\sigma_{min}=1\times10^{-3}$ and $\mathrm{d}l_{s}=5.6\times 10^{-2}$ used throughout the study. 
  }
    \label{fig:conv_mesh}
\end{figure}

As mentioned in $\S$\ref{sec:bc}, the grid involves a region with non-uniform cells encompassing the sphere (purple zone in figure \ref{fig:zones}), and a region where cells are maintained uniform in one direction at larger distances from the body. In the non-uniform region, the cell size is gradually increased as the distance to the sphere surface increases, following a geometric progression. The parameters controlling the grid are thus (i) the size of the cells closest to the sphere and those standing along the rotation axis, i.e. the circumferential length of the cells adjacent to the sphere surface and the radial thickness of the row of cells closest to the sphere and to the rotation axis; (ii) the common ratios of the geometric progressions controlling the variations of the cell size in both directions; (iii) the radial and axial locations at which the transition between the non-uniform and uniform regions takes place; and (iv) the total size of the computational domain in both directions. In what follows, all sizes are expressed in dimensionless form, being normalized by the sphere radius.\\
 \indent As figure \ref{fig:mesh} shows, the grid is singular at the poles of the sphere, which makes the control of the cells located in the pole vicinity of particular importance. To select the thickness $\Delta_{p}$ of the cells adjacent to the sphere surface and closest to the poles (i.e. touching the rotation axis), we first examined the largest values of $\mathcal{R}e$ and $\mathcal{T}a$ we planned to consider, namely $\mathcal{R}e=300$ and $\mathcal{T}a=450$. With these values, the thickness of the `inertial' boundary layer and that of the Ekman layer are close to $0.058$ and $0.047$, respectively. As the present code is known to properly describe the local velocity profiles with $4-5$ cells standing in the boundary layer \citep{Magnaudet2007,Auguste2018}, selecting $\Delta_{p}\approx1\times10^{-2}$ is appropriate. Since the cells thin down along the axis as the distance to the sphere increases, the above value for $\Delta_{p}$ is obtained by selecting a minimum radial cell size $\mathrm{d}\sigma_{min}=1.10^{-3}$ at the upstream and downstream extremities of the non-uniform region, which we fixed at $|z|=50$.  In the sphere vicinity, the cells in a given row are much thinner close to the equator than at the poles. With the above choice for $\mathrm{d}\sigma_{min}$, the cells adjacent to the sphere and closest to the equator are $\approx1.2\times10^{-3}$ thick, which guarantees that the large velocity gradients expected in the equatorial part of the boundary layer are fully captured. We set the common ratio dictating the radial growth of the cells standing in the non-uniform region to $1.10$. The transition between the non-uniform and uniform regions is fixed at $\sigma=30$, so that the radial size of the cells located close to the common boundary of the two regions and beyond it (blue and white regions in figure \ref{fig:zones}) is approximately $2.5$. With these characteristics, $86$ cells are distributed radially across the non-uniform region.
 Regarding the discretization in the polar direction, it was shown by \cite{Auguste2018} that, in a purely inertial flow, a uniform description of the sphere surface with $64$ cells from pole to pole provides converged results at least up to $\mathcal{R}e=500$. Therefore, a slightly less refined discretization is sufficient in the present context, and we selected an angular resolution $\Delta\theta=\pi/56$, which yields cells with length $\mathrm{d}l_{s}\approx5.6\times10^{-2}$ along the sphere surface. Beyond the poles, the first cell along the rotation axis is constrained to have the same length, $\mathrm{d}l_{s}$, as those located along the sphere surface. Then, moving away from the body along the axis, the cells are gradually lengthened following another geometrical progression with a common ratio of $1.05$. $75$ cells are distributed along the axis in the non-uniform region, up to $|z|=50$. At this position, the cells are approximately $2.5$ long and keep the same length beyond that point. Another $54$ uniform cells having this length are distributed along the rotation axis for $|z|>50$ (red region in figure \ref{fig:zones}), so that the computational domain ends at $z=\pm\mathcal{L}=\pm180$.\\
  \indent To choose the outer dimensionless radius $\mathcal{L}_\sigma$ of the domain, it is relevant to consider situations in which inertial effects dominate over those of rotation, as the latter are expected to `tighten' the flow along the rotation axis (apart from the radiation of inertial waves which is specifically handled by the sponge layer). Since the smaller $\mathcal{R}e$ the larger the radial distance over which the sphere-induced disturbance diffuses, we examined the situation corresponding to the minimum Reynolds number to be considered in this study, i.e. $\mathcal{R}e=5$. In this regime, it was shown by \cite{magnaudet_accelerated_1995} that selecting $\mathcal{L}_\sigma=40$ guarantees the absence of spurious confinement effects. This is why, keeping in mind the additional width to be occupied by the sponge layer, we opted for $\mathcal{L}_\sigma=60$, which is achieved by adding $10$ cells with a uniform thickness beyond the outer boundary of the non-uniform region. It may be noticed that \cite{minkov_motion_2000} concluded that the lateral boundary has a negligible effect as soon as $\mathcal{L}_\sigma>5$ for a disc in the low-$\mathcal{R}o$ regime, typically $\mathcal{R}o\leq1\times10^{-2}$. However, this conclusion certainly no longer holds for larger Rossby numbers, typically in the range $0.1\lesssim\mathcal{R}o\lesssim10$, in which most of the computations performed here stand. \\ 
\indent We specifically assessed the influence of $\mathrm{d}l_{s}$, $\mathrm{d}\sigma_{min}$ (hence, $\Delta_{p}$) and $\mathcal{L}_\sigma$ on the case $\mathcal{R}o = 0.1$, $\mathcal{T}a=400$ (i.e. $\mathcal{R}e = 40$) for which the Taylor columns have a moderate elongation upstream and downstream of the sphere. 
Figure \ref{fig:conv_mesh} shows how the drag coefficient varies with these three quantities. 
In all three cases, $C_D$ changes by less than $0.5\%$ in the range within which the parameters are varied. The most sensitive of them turns out to be the minimum radial cell size, $\mathrm{d}\sigma_{min}$. This is no surprise since this parameter controls the spatial resolution available to capture the boundary layer. As the drag varies by only $0.1\%$ in between the smallest two values, we considered that grid convergence is achieved with $\mathrm{d}\sigma_{min}=1\times 10^{-3}$ and retained this value throughout the study. With the above choices for the various grid parameters and the domain size $(\mathcal{L}=180, \mathcal{L}_\sigma=60)$, the total grid involves $2\times(28+75+54)\times(86+10)=314 \times 96$ cells in the axial and radial directions, respectively.\\  
\indent Compared with the grid characteristics described above, the extended domain with $\mathcal{L}=962$ is obtained by increasing the size of the nonuniform region in the axial direction up to 167 sphere radii. Keeping the common ratio and the discretization of the sphere surface unchanged, this is achieved by placing $101$ cells in the nonuniform zone, the largest of which is  approximately $8.75$ sphere radii long. Then, keeping this length unchanged, the domain is extended up to $\mathcal{L}=962$ by adding another $91$ uniform cells. In this case, the grid involves $2\times(28+101+91)\times(86+10)=440 \times 96$ cells in the axial and radial directions, respectively. 

\section{Axial confinement effects} \label{sec:app_conf}

\begin{figure}
    \centering
    \includegraphics[width=0.7\textwidth]{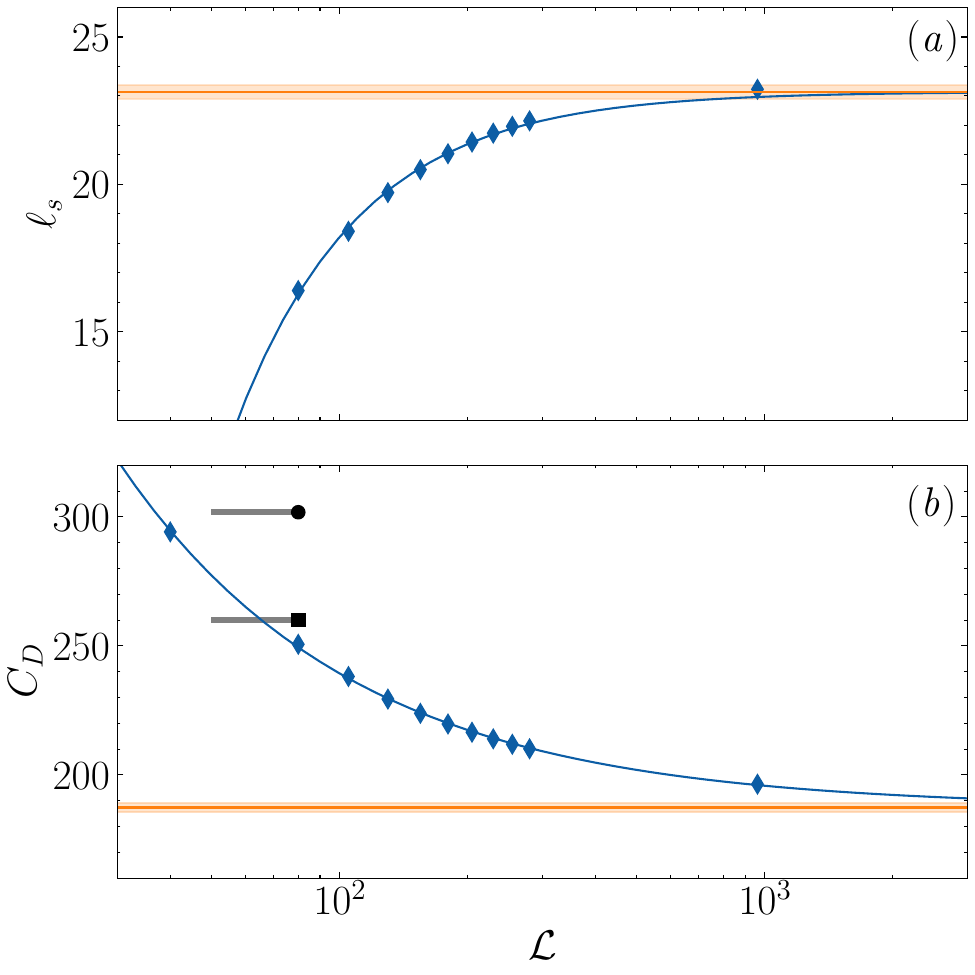}
    \caption{Influence of the axial confinement for $\mathcal{R}o = 0.02$ and $\mathcal{T}a = 445$ ($\mathcal{R}e = 8.9$). $(a)$: length of the upstream recirculation region; $(b)$: drag coefficient. 
    \textcolor{black}{$\blacklozenge$}: present results; \textcolor{orange}{------}: prediction from \citet{tanzosh_motion_1994} at $\mathcal{R}o=0$ in an unbounded domain (the shaded area corresponds to the $\pm 1 \%$ interval around the asymptotic value); \textcolor{black}{------}: best fit of present results, constrained to tend to the asymptotic value for $\mathcal{L} \to \infty$; $\bullet$: drag determined by \citet{maxworthy_flow_1970} in a container with $\mathcal{L} = 80$ (interpolated from data at $\mathcal{R}e = 7.8$ and $10.4$, both with $\mathcal{T}a=445$); $\blacksquare$: extrapolation of the same experimental data based on \eqref{eq:max_extra}. The grey bar indicates the range of $\mathcal{L}$ spanned by the particules used in the experiments as they rose along the `viewing box' within which the drag was determined. 
    }
    \label{fig:lus_vs_lx}
\end{figure}
\begin{figure}
    \centering
    \includegraphics[width=0.7\textwidth]{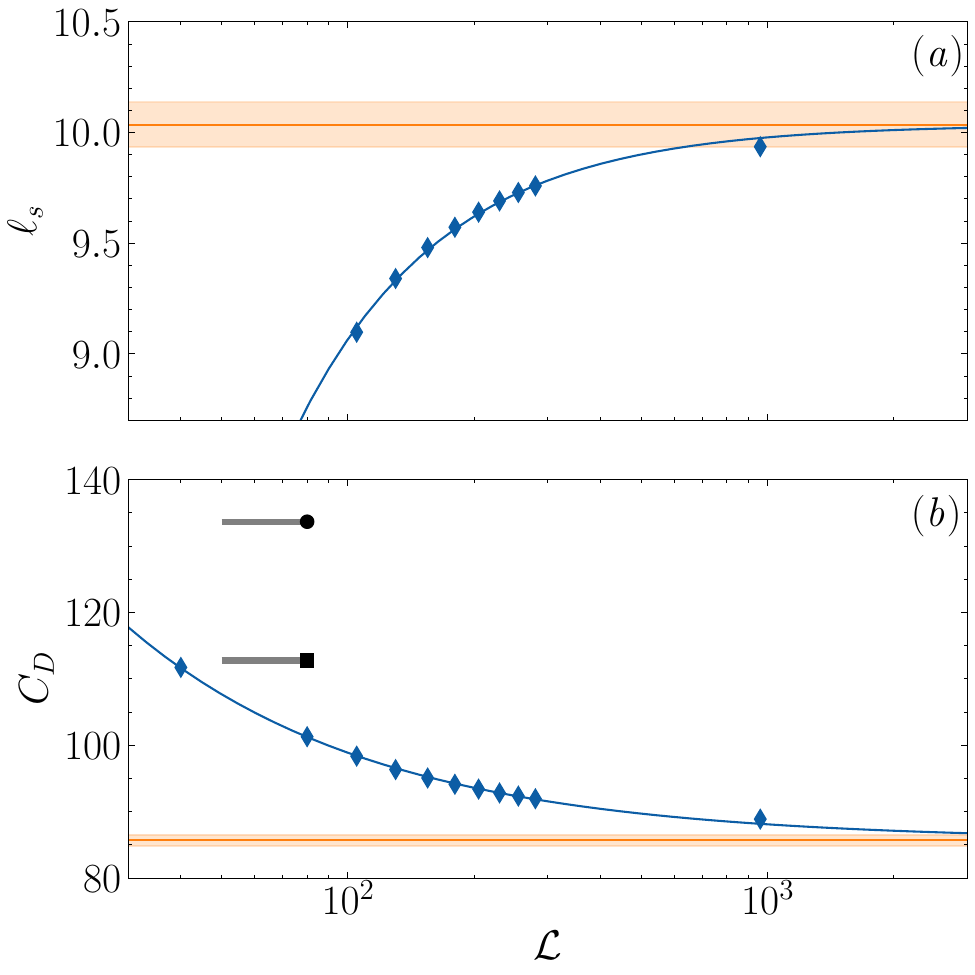}
    \caption{Same as figure \ref{fig:lus_vs_lx} for $\mathcal{R}o = 0.046$ and $\mathcal{T}a = 193$ (i.e. still $\mathcal{R}e = 8.9$).  
    }
    \label{fig:lus_vs_lx_t193}
\end{figure}

In a preliminary simulation with a short domain ($\mathcal{L} \approx 40$), we noticed that, with $\mathcal{R}e=8.9$ and $\mathcal{T}a = 445$ ($\mathcal{R}o=0.02$), the drag deviated from (\ref{eq:TS}) by approximately $50 \%$. Examining the flow revealed that, although the upstream and downstream recirculation regions extended only over $15 \%$ of the available length,  the Taylor column reached both the inlet of the domain and the downstream sponge region. For this reason, the axial velocity abruptly recovered its prescribed value when approaching the end walls. The inescapable conclusion was that an upper or lower boundary located `too close' to the sphere compresses the Taylor column and may induce a large artificial drag increase.

To determine the minimum axial size of the domain beyond which confinement effects become negligible (or rather `acceptable'), we performed a detailed sensitivity study with the above $(\mathcal{R}e, \mathcal{T}a)$ set. More specifically, we built a series of grids of increasing length, varying $\mathcal{L}$ over more than one order of magnitude, from $\mathcal{L}=40$ to $\mathcal{L}=962$.  The sphere was kept halfway between the two end walls in all cases.  
Confinement effects were evaluated by comparing the drag coefficient with \eqref{eq:TS}, and the length $\ell_{s}$ of the upstream recirculation region with the numerical prediction of \cite{tanzosh_motion_1994} in the zero-$\mathcal{R}o$ limit, $\ell_{s} = 0.052\, \mathcal{T}a$ (see \S\,\ref{sec:recirc}), respectively.
Results of this study are reported in figure \ref{fig:lus_vs_lx}. The length of the upstream recirculation region (figure \ref{fig:lus_vs_lx}$(a)$) is observed to be still significantly under-predicted with the standard domain length $\mathcal{L}=180$. In order to agree within $1 \%$ with the zero-$\mathcal{R}o$ value, a minimum half-length $\mathcal{L}\approx700$ is required. For sufficiently short domains ($\mathcal{L}\lesssim30$), no upstream recirculation is detected any more, the axial velocity never changing sign upstream of the sphere. The situation is even more dramatic regarding the drag, as figure \ref{fig:lus_vs_lx}$(b)$ shows: extrapolating the results obtained with domain half-lengths up to $10^3$, one has to conclude that it is only beyond $\mathcal{L}\approx 10^4$ that the drag may agree within one percent with the theoretical prediction.
In figure \ref{fig:lus_vs_lx}$(b)$ we added the drag determined by \citet{maxworthy_flow_1970} for the same set of parameters (bullet). The discrepancy with the theoretical prediction is roughly $55 \%$. The `corrected' results based on the extrapolation \eqref{eq:max_extra} (black square) still overestimates $C_D$ by about $35 \%$. 
 \\
\indent We repeated the analysis with a smaller value of the Taylor number, $\mathcal{T}a=193$, still with $\mathcal{R}e = 8.9$ ($\mathcal{R}o=0.046$). The results are presented in figure \ref{fig:lus_vs_lx_t193}. In this case, the upstream recirculation is significantly smaller ($\ell_s\approx10$). However, figure \ref{fig:lus_vs_lx_t193}$(a)$ indicates that the computational domain has to be even longer than in the previous case ($\mathcal{L}\approx 10^3$) for the numerical estimate of $\ell_s$ to agree within one percent with the zero-$\mathcal{R}o$ prediction, and the drag coefficient agrees within one percent with \eqref{eq:TS} only for domain lengths beyond $\approx 2\times10^3$. Note, however, that $\mathcal{R}o$ being larger than in the previous case, the zero-$\mathcal{R}o$ prediction of \cite{tanzosh_motion_1994} is expected to be slightly less accurate. Therefore, there is no guarantee that a $1\%$ agreement with these predictions is to be expected, even on an infinitely long domain. \\
\begin{figure}
    \centering
    \includegraphics[width=0.7\textwidth]{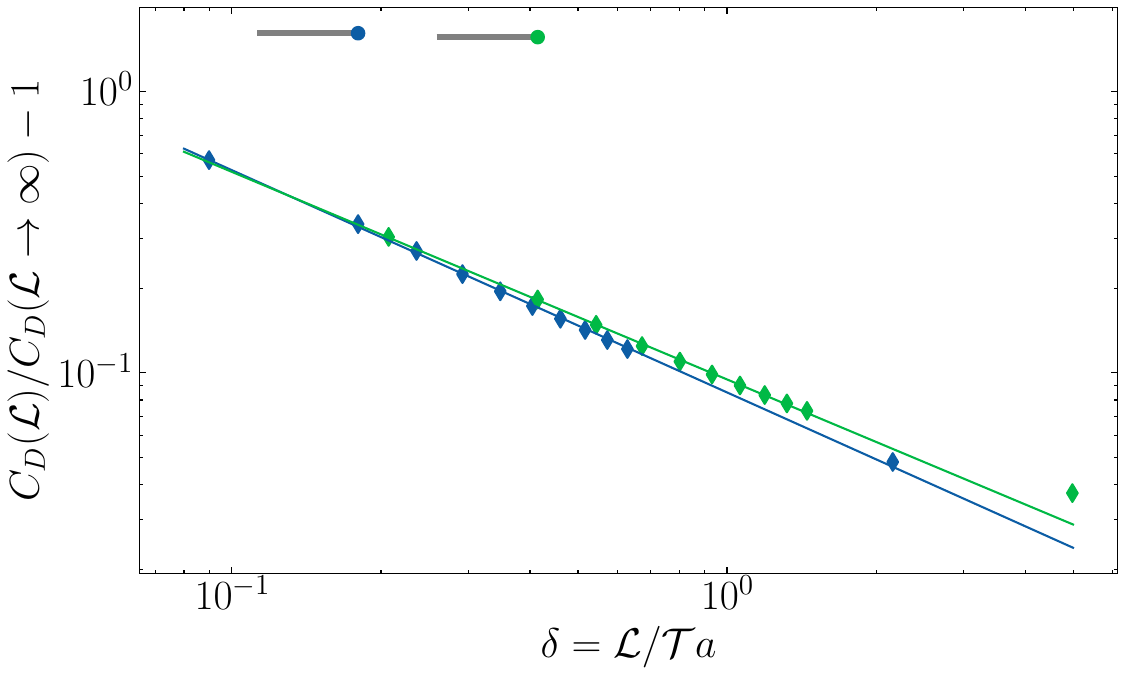}
    \caption{Variations of the drag coefficient with the normalized domain half-size $\delta=\mathcal{L}/\mathcal{T}a$.  \textcolor{green}{$\blacklozenge$}: $\mathcal{R}o = 0.046$ , $\mathcal{T}a = 193$;  \textcolor{black}{$\blacklozenge$}: $\mathcal{R}o = 0.02$, $\mathcal{T}a = 445$. Bullets and grey lines refer to Maxworthy's experimental conditions. Solid line: correlation \eqref{fitconf}.}
    \label{Cd_delta}
\end{figure}
\indent The above results are replotted in figure \ref{Cd_delta}, with the domain half-length rescaled by the Taylor number, following the asymptotic analysis of \cite{Hocking1979}. The two sets of results are seen to follow a power law, with slightly $\mathcal{R}o$-dependent parameters. These results are accurately fitted by the empirical formula
\begin{equation}
\frac{C_D(\mathcal{L})}{C_D(\mathcal{L}\rightarrow\infty)}-1\approx7.78\times10^{-2}(1+4.7\mathcal{R}o)\delta^{-0.83+2.0\mathcal{R}o}\,.
\label{fitconf}
\end{equation}
Of course, the $\mathcal{R}o$-dependent correction has a limited range of validity that does not presumably extend beyond $\mathcal{R}o\approx0.1$, being based on only two low-$\mathcal{R}o$ data sets. Moreover, the sensitivity to $\mathcal{R}o$ may be artificial, since our evaluation of the confinement effect is based on the difference with the zero-$\mathcal{R}o$ prediction \eqref{eq:TS}, the accuracy of which is expected to decrease as $\mathcal{R}o$ increases. Nevertheless, this fit might be useful to obtain a rough estimate of axial confinement effects in future experiments. Of course, the differences between the no-slip conditions applying to closed containers and the boundary conditions used on the two end surfaces (plus the presence of the sponge layer) in the present simulations must be kept in mind. Also, the fact that the sphere is held fixed midway between the end surfaces in the simulations while it moves toward one of them and away from the other in experiments makes a significant difference.\\
\indent Overall, it turns out that the axial length of the computational domain, or equivalently the height of the experimental container, is critical in the present problem, owing to the direct kinematic interaction of the Taylor column with the end walls in the low-$\mathcal{R}o$ large-$\mathcal{T}a$ regime. Axial confinement effects appear as the main source of discrepancy between experimental data and the prediction \eqref{eq:TS} for the drag in this regime. Consequently, in \S\,\ref{loads} we only discuss results that are almost free of these effects. In practice, we disregarded results obtained in simulations where the axial velocity in the sphere's wake has not relaxed to at least $0.9\, U_{\infty}$ before entering the sponge region. This led us to exclude results belonging to the range $(\mathcal{T}a > 150,\,\mathcal{R}o < 0.125)$ obtained on the standard domain with $\mathcal{L}=180$, and to replace them with results obtained on the extended domain with $\mathcal{L}=962$. 

\bibliographystyle{jfm}
\bibliography{rotating_sphere}

\end{document}